\documentclass[twocolumn]{aastex63}
\usepackage{multirow}
\usepackage[figuresright]{rotating}
\usepackage{subfigure}
\usepackage{epic,eepic}
\usepackage{graphicx}
\usepackage{longtable}
\usepackage{float}
\usepackage{pifont}
\usepackage{gensymb}

\shorttitle{Binaries with possible compact components}
\shortauthors{Li et al.}

\begin{document}

\title{Binaries with possible compact components discovered from the LAMOST Time-Domain Survey of four $K$2 plates}

\correspondingauthor{Song Wang}
\email{songw@bao.ac.cn}

\author{Xue Li}
\affiliation{Key Laboratory of Optical Astronomy, National Astronomical Observatories, Chinese Academy of Sciences, Beijing 100101, China}
\affiliation{College of Astronomy and Space Sciences, University of Chinese Academy of Sciences, Beijing 100049, China}

\author{Song Wang}
\affiliation{Key Laboratory of Optical Astronomy, National Astronomical Observatories, Chinese Academy of Sciences, Beijing 100101, China}

\author{Xinlin Zhao}
\affiliation{Key Laboratory of Optical Astronomy, National Astronomical Observatories, Chinese Academy of Sciences, Beijing 100101, China}
\affiliation{College of Astronomy and Space Sciences, University of Chinese Academy of Sciences, Beijing 100049, China}

\author{Zhongrui Bai}
\affiliation{Key Laboratory of Optical Astronomy, National Astronomical Observatories, Chinese Academy of Sciences, Beijing 100101, China}

\author{Hailong Yuan}
\affiliation{Key Laboratory of Optical Astronomy, National Astronomical Observatories, Chinese Academy of Sciences, Beijing 100101, China}

\author{Haotong Zhang}
\affiliation{Key Laboratory of Optical Astronomy, National Astronomical Observatories, Chinese Academy of Sciences, Beijing 100101, China}

\author{Jifeng Liu}
\affiliation{Key Laboratory of Optical Astronomy, National Astronomical Observatories, Chinese Academy of Sciences, Beijing 100101, China}
\affiliation{College of Astronomy and Space Sciences, University of Chinese Academy of Sciences, Beijing 100049, China}
\affiliation{WHU-NAOC Joint Center for Astronomy, Wuhan University, Wuhan, Hubei 430072, China}

\let\cleardoublepage\clearpage
\begin{abstract}
Time-domain (TD) spectroscopic data from the Large Sky Area Multi-Object Fiber Spectroscopic Telescope (LAMOST) can provide accurate and high-cadence radial velocities (RVs). In this work, we search for binaries with compact components with RV monitoring method by using the LAMOST TD survey of four $K$2 plates. Three binary systems including an unseen white dwarf or neutron star are found. For each binary system, we estimate the stellar parameters of the visible star and orbital parameters, and finally calculate the binary mass function and the minimum mass of the unseen star. 
No obvious double-lined feature is seen from the LAMOST medium-resolution spectra of the three sources.
In addition, we found no X-ray counterpart for all these sources but UV companions for two of them.
Spectral disentangling also shows no additional component with optical absorption spectra, supporting that these systems contain compact objects.
\end{abstract}

\keywords{binaries: general --- white dwarfs --- stars: neutron}

\section{INTRODUCTION}
\label{intro.sec}

Most stars end their lives as compact objects, including white dwarfs, neutron stars, and black holes. The evolution tracks are determined by stellar initial masses, metallicities, and rotational velocities, etc.
In the multi-messenger era, many methods (e.g., gravitational wave, microlensing, and X-ray) are utilized to search for single compact objects or compact objects in binaries.
Radial velocity (RV) monitoring has been proved to be a feasible method in discovering unseen massive compact companions in binary systems with the help of large spectroscopic surveys. 
In recent years, more than 10 X-ray quiescent stellar-mass black holes have been discovered through RV monitoring \citep[e.g.,][]{2014Natur.505..378C,2019Sci...366..637T,2019Natur.575..618L,2020A&A...637L...3R,2021arXiv210102212J}, although the nature of some ones are still under debate \citep[][]{2020A&A...639L...6S,2020MNRAS.493L..22E,2020Natur.580E..11A}.
Some neutron stars \citep{2021ApJ...909..185S} or white dwarfs \citep{2021MNRAS.505.2051E} in binary systems were also discovered through this method.

As one of the most powerful optical spectrum survey telescope ($\sim$4000 fibers),
the Large Sky Area Multi-Object Fiber Spectroscopic Telescope (hereafter LAMOST, also called the GuoShouJing Telescope) 
contributes more than one million spectra every year.
It is a reflecting Schmidt telescope, with an effective aperture of 4 m and a field of view of 5 degrees.
\citep{2012RAA....12.1197C,2012RAA....12..723Z}. 
LAMOST has started a second 5-year sky survey from 2018, containing both time domain (TD) and non-TD surveys. It performs both low-resolution spectral (LRS) and medium-resolution spectral (MRS) observations with $\Delta \lambda / \lambda \sim$ 1800 and $\sim$ 7500, respectively \citep{2020arXiv200507210L}. 
The LRS observations cover the wavelength range of 3650--9000 \AA, while the MRS observations provide spectra covering the wavelength range from 4950 \AA \ to 5930 \AA \ for the blue arm and from 6300 \AA \ to 6800 \AA \ for the red arm, respectively.
The LAMOST TD survey will monitor about 200,000 stars with an average of 60 MRS exposures in five years,
with a limiting magnitude around $G$ $\approx$ 15 mag \citep{2020arXiv200507210L}. 

From 2019, we performed a TD survey of four K2 plates using the LAMOST telescope \citep{2021RAA....21..292W}. 
In the first year, the LRS survey gained about 767,000 spectra over 282 exposures during 25 nights, and the MRS survey derived about 478,000 spectra with 177 exposures over 27 nights. More than 70\%/50\% of the low-resolution/medium-resolution spectra have signal-to-noise ratio ({\it SNR}) higher than 10. 
We selected  binaries with possible compact components from these observations and performed detailed analysis by using multi-band data.
This paper is organized as follows.
In Section \ref{obs.sec}, we present the spectroscopic observation and data reduction. Section \ref{star.sec} shows the basic information of the visible star, including atmospheric parameters, distances, and masses, etc. In Section \ref{orbit.sec}, we perform RV fitting using {\it The Joker} code. Finally, we give the discussion and summary in Section \ref{diss.sec} and \ref{sum.sec}.

\section{Source selection and data reduction}
\label{obs.sec}

\subsection{Source selection}
\begin{figure*}
    \center
    \includegraphics[width=0.98\textwidth]{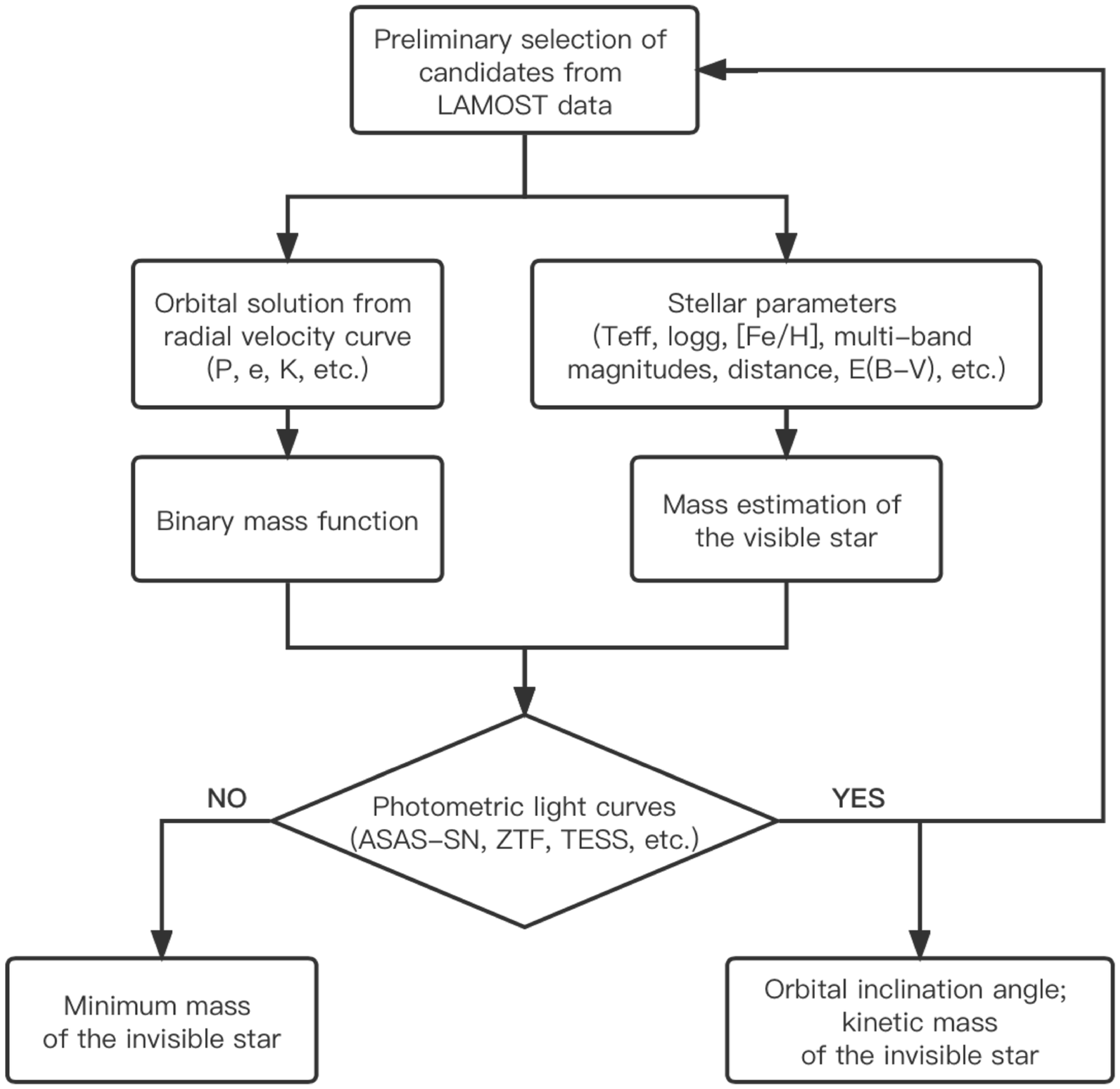}
    \caption{Summary flowchart of the source selection process in this paper.}
    \label{flowchart.fig}
\end{figure*}

We selected candidate binaries with compact objects from our TD spectroscopic survey of four $K$2 plates \citep{2021RAA....21..292W}. 
Figure \ref{flowchart.fig} summarizes the steps in a flowchart for reference.
First, sources with clear RV variation ($\Delta$RV $>$ 10 km/s) were selected based on the LAMOST data. Double-lined spectroscopic binaries were excluded.
Second, we derived the orbital solution (e.g., period $P$, eccentricity $e$, and semi-amplitude $K$) by fitting the RV data using {\it The Joker} code \citep{2017ApJ...837...20P}. 
The binary mass function was calculated with the orbital parameters. 
Third, we calculated the evolutionary mass and spectroscopic mass of the visible star basing on their stellar parameters. 
Fourth, in order to constrain the orbital inclination angle, we collected light curves (LCs) from some wide-field photometric surveys, including the Catalina Sky Survey \citep[CSS;][]{2009ApJ...696..870D}, All-Sky Automated Survey for Supernovae \citep[ASAS-SN;][]{2017PASP..129j4502K}, Zwicky Transient Facility \citep[ZTF;][]{2019PASP..131a8002B}, and $K$2. The LCs can also help derive orbital period and do candidate selection (e.g., excluding EA-type normal binary).
Finally, the mass of invisible star was calculated by using the orbital parameters.
If the (minimum) mass of the invisible star is close to the visible star or larger than 1 M$_{\odot}$, they will be picked out for detailed analysis.

Following above steps, we found three single-lined spectroscopic binaries showing periodic RV variation: G6084, G6081 and G6405. Their basic information are shown in Table \ref{basic.tab}.

\begin{table*}
\caption{Basic information of G6084, G6081, and G6405.}\label{basic.tab}
\centering
\setlength{\tabcolsep}{2.5pt}
 \begin{tabular}{cccc}
\hline\noalign{\smallskip}
Parameters & G6084 & G6081 & G6405 \\
\hline\noalign{\smallskip}
GAIA eDR3 source ID & 608426858154004864 & 608189290627289856 & 64055043471903616\\
RA(\degree) & 131.96675 & 133.26916 & 57.50692\\
DEC(\degree) & 13.45496 & 13.34231 & 22.31626\\
$N_{\rm LRS}$ & 118 & 113 & 97\\
$N_{\rm MRS}$ & 86 & 106 & 59\\
\noalign{\smallskip}\hline
\end{tabular}
\smallskip\\
$N_{\rm LRS}$ and $N_{\rm MRS}$ are the numbers of low- and medium-resolution spectra obtained by LAMOST, respectively.
\end{table*}

\subsection{LAMOST observations and data reduction}
We collected all LAMOST archive observations for our candidate sources.
The numbers of the low- and medium-resolution exposures are given in Table \ref{basic.tab}.
The LAMOST 2D pipeline includes bias and dark subtraction, flat field correction, spectrum extraction, sky background subtraction, and wavelength calibration, etc \citep[see][for details]{2015RAA....15.1095L}.
The released spectra are in the vacuum wavelength scale and have been converted to the heliocentric frame of reference.

We used the cross-correlation function to calculate RV values from the MRS and LRS data with {\it SNR} higher than 5.
We compared our results with the released RVs from LAMOST  catalogs\footnote{http://www.lamost.org/dr9/v1.0/catalogue}, and found that our results are more accurate. 
The RV uncertainty is calculated as the square root of the sum of the squares of the measurement error and the wavelength calibration error. 

\begin{figure*}
    \center
    \includegraphics[width=\textwidth]{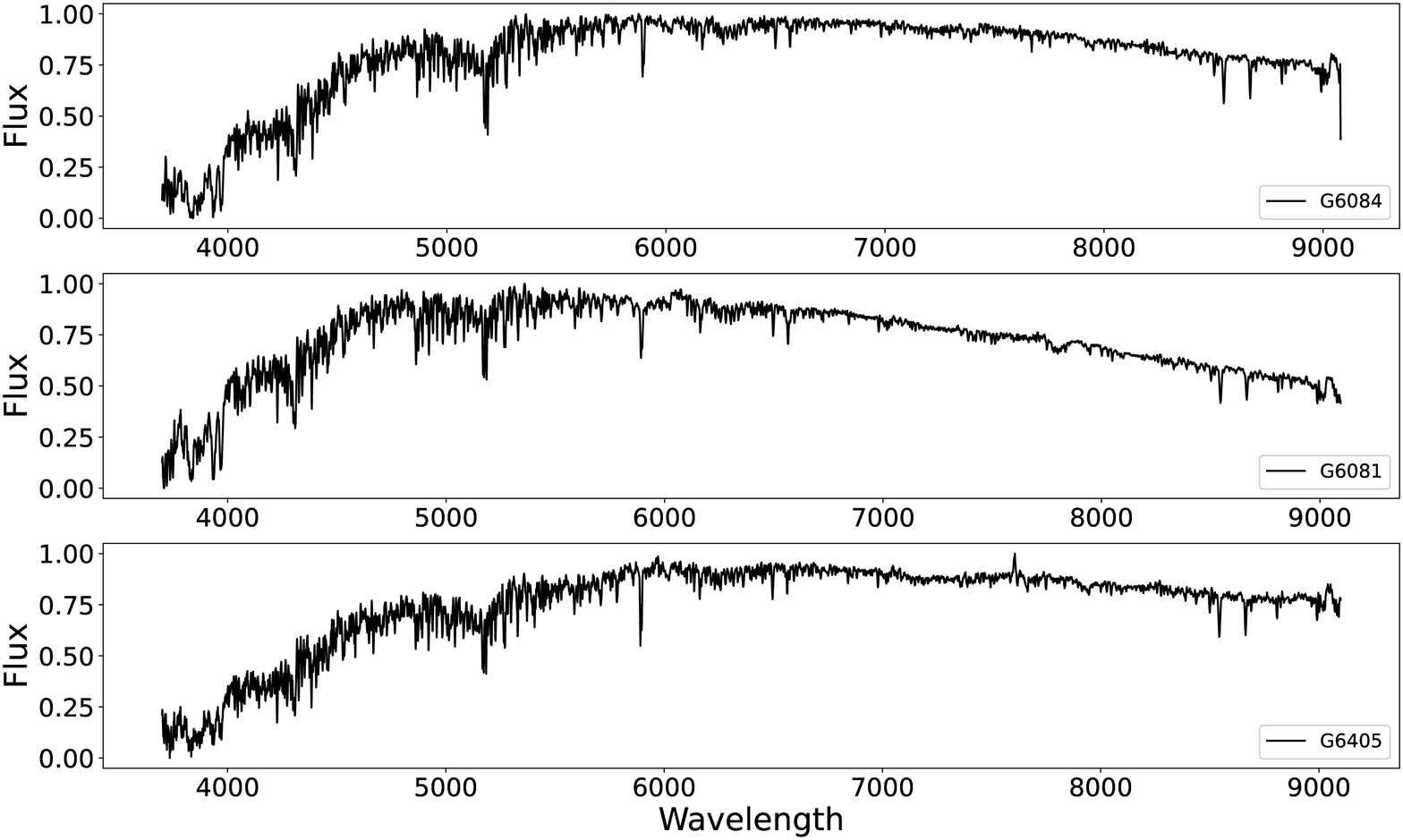}
    \caption{Low-resolution spectra of G6084 (top panel), G6081 (middle panel), and G6405 (bottom panel) from LAMOST observations.}
    \label{lowspecs.fig}
\end{figure*}

\section{The visible star}
\label{star.sec}

\subsection{Stellar parameters}
\label{para.sec}

Our targets were observed at multiple epochs, thus we derived {\it SNR}-weighted average values and corresponding errors for the stellar parameters following \citep{2020ApJS..251...15Z}:
\begin{equation} \label{eqweight}
\overline{P} = \frac{\sum_k w_k \cdot P_{k}}{\sum_k w_k}
\end{equation}
and
\begin{equation}
\sigma_w(\overline{P}) = \sqrt{\frac{N}{N-1}\frac{\sum_k w_k \cdot (P_{k} - \overline{P})^2}{\sum_k w_k}}.
\end{equation}
The index $k$ is the epoch of the measurements of parameter $P$ (i.e., $T_{\rm eff}$, log$g$, and [Fe/H]) for each star, and the weight $w_k$ is estimated with the square of the {\it SNR} of each spectrum. 

To G6084, by using the LAMOST DR8 low-resolution catalog, the stellar parameters were calculated as $T_{\rm eff} = 4737{\pm 24}$ K, log$g$ $=3.20{\pm 0.05}$ and [Fe/H] $=-0.37{\pm 0.01}$.
By using the DR8 medium-resolution catalog, the parameters are $T_{\rm eff} = 4652{\pm 20}$ K, log$g$ $=3.16{\pm 0.07}$, and [Fe/H] $=-0.49{\pm 0.06}$.
\cite{2021RAA....21..292W} has listed stellar parameters estimated by different methods (i.e., LASP, DD-Payne, and SLAM). Comparisons with those results showed that the parameters from LAMOST DR8 low-resolution catalog are more reliable.
We derived the extinction Bayestar19\footnote{http://argonaut.skymaps.info/usage} $\approx$ 0.01 from the Pan-STARRS DR1 dust map \citep{2015ApJ...810...25G} corresponding to a distance of approximately 1100 pc from {\it Gaia} DR2 \citep{2018AJ....156...58B}.
The $E(B-V)$ value was calculated with $E(B-V) =0.884 \times {\rm (Bayestar19)}$. Table \ref{parameters.tab} lists the parameters in detail.

There is a bright source close to G6081, and some spectroscopic observations of G6081 were contaminated by the bright source. By excluding these observations, we derived the atmospheric parameters from LAMOST DR8 low-resolution catalog: $T_{\rm eff} = 5320{\pm 160}$ K, log$g$ $=4.33{\pm 0.16}$, and [Fe/H] $=-0.30{\pm 0.06}$.
The distance is $\approx$320 pc and the extinction is $E(B-V)\approx0.009$. 

\begin{figure}
   \center
   \includegraphics[width=0.52\textwidth]{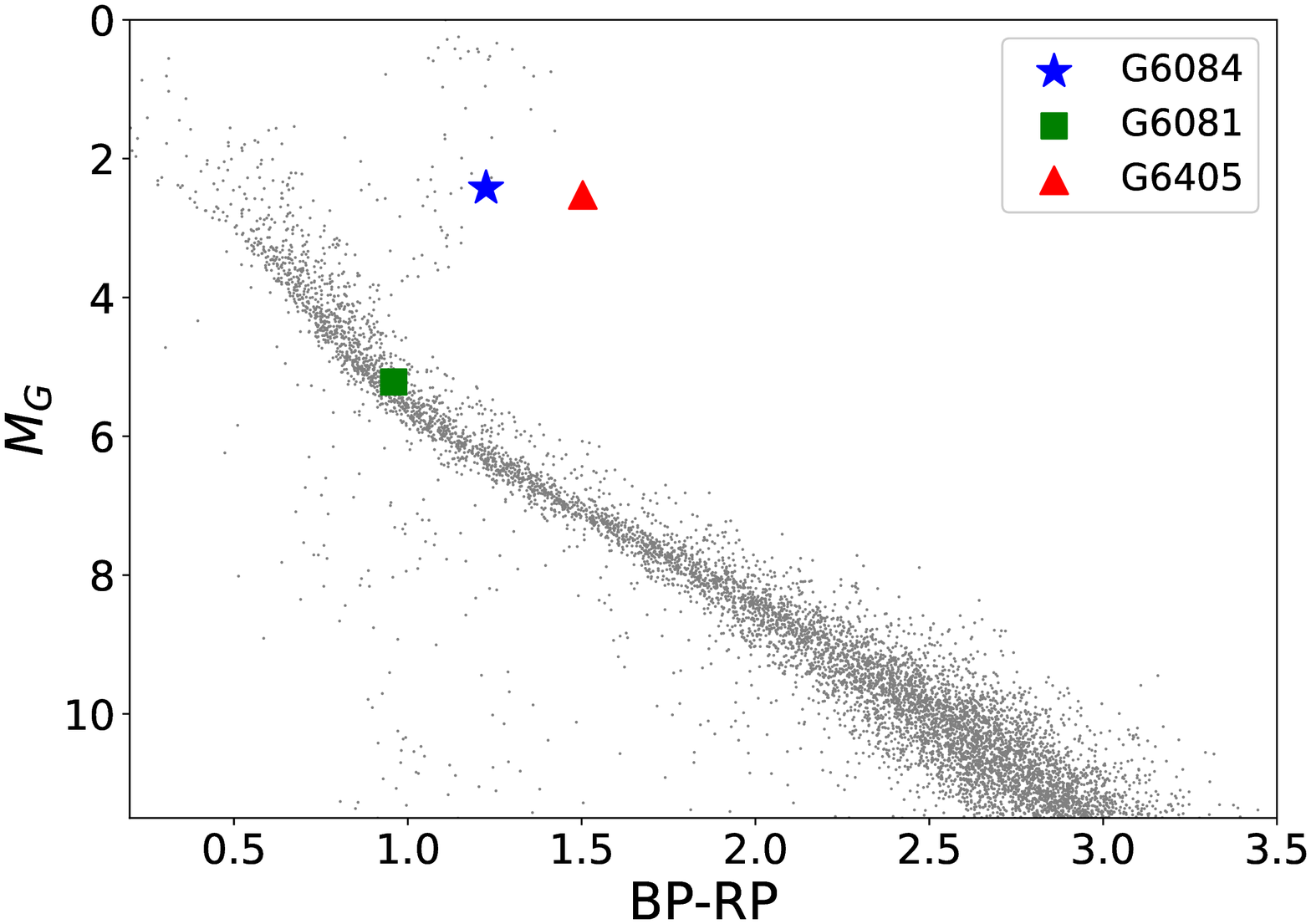}
   \caption{Hertzsprung–Russell diagram of G6084 (blue asterisk), G6081 (green square), and G6405 (red triangle). The grey points are the stars from {\it Gaia} eDR3 with distances $d <$ 150 pc, $G_{\rm mag}$ between 4--18 mag, and Galactic latitude |$b$| $>$ 10. No extinction was corrected for these stars.}
   \label{hrdiagram.fig}
\end{figure}

For G6405, from LAMOST DR8 low-resolution catalog, we got the atmospheric parameters of $T_{\rm eff} = 4709{\pm 34}$ K, log$g$ $=3.01{\pm 0.07}$, and [Fe/H] $=-0.09{\pm 0.06}$. 
The parameters estimated from different methods, collected from \cite{2021RAA....21..292W}, are a little different (Table \ref{G6405pars.tab}).
In this paper, we used the parameters from the DR8 low-resolution catalog. 
The distance is about 2 kpc and the extinction is $E(B-V)\approx0.17$. 
The absolute $G$-band magnitudes of G6084, G6081, and G6405 are 2.42 mag, 5.22 mag, and 2.52 mag, respectively.
Figure \ref{hrdiagram.fig} shows the positions of these targets in the Hertzsprung–Russell diagram.

\begin{table}
\caption{Stellar parameters of G6084, G6081, and G6405.\label{parameters.tab}}
\centering
\setlength{\tabcolsep}{2pt}
 \begin{tabular}{cccc}
\hline\noalign{\smallskip}
Parameters & G6084 & G6081 & G6405 \\
\hline\noalign{\smallskip}
$T_{\rm eff}$ (K) & $4737{\pm 24}$ & $(532{\pm 16})\times10$ & $4709{\pm 34}$\\
$log$g & $3.20{\pm 0.05}$ & $4.33{\pm 0.16}$ & $3.01{\pm 0.07}$\\
${\rm [Fe/H]}$ & $-0.37{\pm 0.01}$ & $-0.30{\pm 0.06}$ & $-0.09{\pm 0.06}$\\
\hline
$d$ (pc) & $1101^{+48}_{-44}$ & $320^{+5}_{-5}$ & $(200^{+19}_{-16})\times10$\\
$\varpi$ (mas) & $0.88{\pm 0.04}$ & $3.10{\pm 0.05}$ & $0.46{\pm 0.04}$\\
$E(B-V)$ & 0.009 & 0.009 & 0.17 \\
\hline
$G$ (mag) & $12.66{\pm 0.002}$ & $12.78{\pm 0.003}$ & $14.51{\pm 0.002}$\\
$BP$ (mag)  & $13.23{\pm 0.008}$ & $13.19{\pm 0.009}$ & $15.21{\pm 0.007}$\\
$RP$ (mag)  & $11.98{\pm 0.007}$ & $12.22{\pm 0.007}$ & $13.69{\pm 0.005}$\\
$J$ (mag)  & $11.10{\pm 0.023}$ & $11.49{\pm 0.021}$ & $12.48{\pm 0.023}$\\
$H$ (mag)  & $10.54{\pm 0.025}$ & $11.04{\pm 0.019}$ & $11.90{\pm 0.022}$\\
$K_{\rm S}$ (mag)  & $10.45{\pm 0.024}$ & $11.00{\pm 0.020}$ & $11.74{\pm 0.018}$\\
\noalign{\smallskip}\hline
\end{tabular}
\end{table}

\begin{table}
\caption{Atmospheric parameters of G6405 estimated from different methods \label{G6405pars.tab}}
\centering
\setlength{\tabcolsep}{1.5pt}
 \begin{tabular}{ccccc}
\hline\noalign{\smallskip}
Obs. &Methods & $T_{\rm eff}$ (K) & log$g$ (cm/s$^2$) & [Fe/H] \\
\hline\noalign{\smallskip}
 \multirow{3}*{LRS} & DR8    & $4709{\pm 34}$ & $3.01{\pm 0.07}$ & $-0.09{\pm 0.06}$\\
&  LASP  & $4703{\pm 66}$ & $2.95{\pm 0.14}$ & $-0.13{\pm 0.07}$\\
& DD-payne  & $4731{\pm 30}$ & $3.27{\pm 0.07}$ & $-0.28{\pm 0.06}$\\
\hline
 \multirow{3}*{MRS} & DR8  & $4670{\pm 91}$ & $2.92{\pm 0.31}$ & $-0.64{\pm 0.21}$\\
 & LASP & $4579{\pm 34}$ & $2.60{\pm 0.30}$ & $-0.69{\pm 0.03}$\\
& SLAM  & $4962{\pm 126}$ & $3.30{\pm 0.24}$ & $-0.12{\pm 0.07}$\\
\noalign{\smallskip}\hline
\end{tabular}
\end{table}

\subsection{Mass Determination}
\label{pmass.sec}

We calculated the stellar masses by two methods: one is to use the stellar evolution model, and the other is to use the observed photometric and spectroscopic parameters.

\subsubsection{Evolutionary Mass Estimation}
\label{subsubsect:mass_evo}

The {\it isochrones} Python module \citep{2015ascl.soft03010M} can be used to fit stellar models to photometric or spectroscopic parameters.
The input includes the measured effective temperature, surface gravity, metallicity, multi-band magnitudes ($G$, $G_{\rm BP}$, $G_{\rm RP}$, $J$, $H$, and $K_{\rm S}$), {\it Gaia} DR2 parallax \citep{2018AA...616A...1G} and extinction $A_V$ ($= 3.1 \times E(B-V)$).
The output is physical or photometric properties provided by the best-fit models \citep{2015ApJ...809...25M}.

The fitted evolutionary masses are $1.17^{+0.10}_{-0.08}$ M$_{\odot}$ for G6084, $0.83^{+0.01}_{-0.01}$ M$_{\odot}$ for G6081, and $1.00^{+0.08}_{-0.06}$ M$_{\odot}$ for G6405, respectively.
The fitting results are displayed in Figure \ref{8503isochrones.fig}, \ref{8050isochrones.fig}, and \ref{128isochrones.fig} in Appendix \ref{iso_appendix.sec}.

\subsubsection{Gravitational mass estimation}
\label{subsubsect:mass_sp}

The gravitational masses were calculated based on the observed spectrosopic and photometric parameters. 
Firstly, we derived the bolometric magnitudes following
\begin{equation}
\label{mbol.eq}
m_{\rm \lambda} - M_{\rm bol} = 5{\rm log}d - 5 + A_{\rm \lambda} - BC_{\rm \lambda}.
\end{equation}
Here, $m_{\rm \lambda}$ is the apparent magnitude of six bands ($G$, $G_{\rm BP}$, $G_{\rm RP}$, $J$, $H$, and $K_{\rm S}$), $d$ is the distance from the {\it Gaia} DR2, and $A_{\rm \lambda}$ is the extinction of each band calculated from the Baystar19 value.
BC is the bolometric correction calculated from the PARSEC database\footnote{http://stev.oapd.inaf.it/YBC/} \citep{2019AA...632A.105C}, with the input of $T_{\rm eff}$, log$g$, and [Fe/H] values. 
Secondary, the bolometric luminosity and the stellar radius can be calculated with the formulae: 
\begin{equation}
    M_{\rm bol} - M_{\odot}  = -2.5{\rm log}\frac{L_{\rm bol}}{L_{\odot}}
\end{equation}
and
\begin{equation}
\label{luminosity.eq}
    L_{\rm bol} = 4\pi~{R}^{2}\sigma~T^{4}.
\end{equation}
Here $M_{\odot}$ is solar bolometric magnitude (4.74 mag) and  $L_{\odot}$ is solar bolometric luminosity (3.828$\times10^{33}$ erg/s).
Finally, we estimated the stellar mass as follows,
\begin{equation}
\label{visible_mass.eq}
    M = \frac{g{R}^{2}}{G}.
\end{equation}

The final masses and corresponding uncertainties, which are the average values and standard deviations of the masses derived from different bands, are:
$1.18{\pm 0.07}$ M$_{\odot}$ for G6084, $0.69{\pm 0.11}$ M$_{\odot}$ for G6081, and $0.74{\pm 0.08}$ M$_{\odot}$ for G6405.

There are some difference between the gravitational and evolutionary masses for G6405, which is mainly caused by the measurement accuracy of stellar parameters.
The relative uncertainty of the distance of G6405 is large and close to 0.2. Using the uncertainty, the gravitational mass of G6405 will be 0.9 (0.63) M$_{\odot}$ if $d$ is adopted as 2190 (1840) pc. In addition, the gravitational mass is sensitive to log$g$, and a small variation of log$g$ will lead to a different mass estimation. The log$g$ value of G6405 is $3.01\pm0.07$, and the mass will be 0.88 M$_{\odot}$ if log$g$ is adopted as 3.08 (similar to the {\it isochrones} fit value). More accurate measurements of these stellar parameters in the future will help resolve the problem.

\subsubsection{Spectral energy distribution fitting}
\label{subsubsect:mass_sed}

We also used the spectral energy distribution (SED) fitting method to estimate the stellar mass with astroARIADNE\footnote{https://github.com/jvines/astroARIADNE}. 
AstroARIADNE is designed to automatically fit broadband photometry to different stellar atmosphere models using Nested Sampling algorithms. 
The {\it Gaia} DR2 ($G$, $G_{\rm BP}$, and $G_{\rm RP}$), 2MASS ($J$, $H$, and $K_{\rm S}$), APASS ($B$, $V$, $g$, $r$, and $i$) and WISE ($W$1 and $W$2) magnitudes were used in the fitting.
The stellar evolutionary models include Phoenix, BTSettl, Castelli \& Kurucz and Kurucz 1993.

\begin{figure}
    \center
    \includegraphics[width=0.48\textwidth]{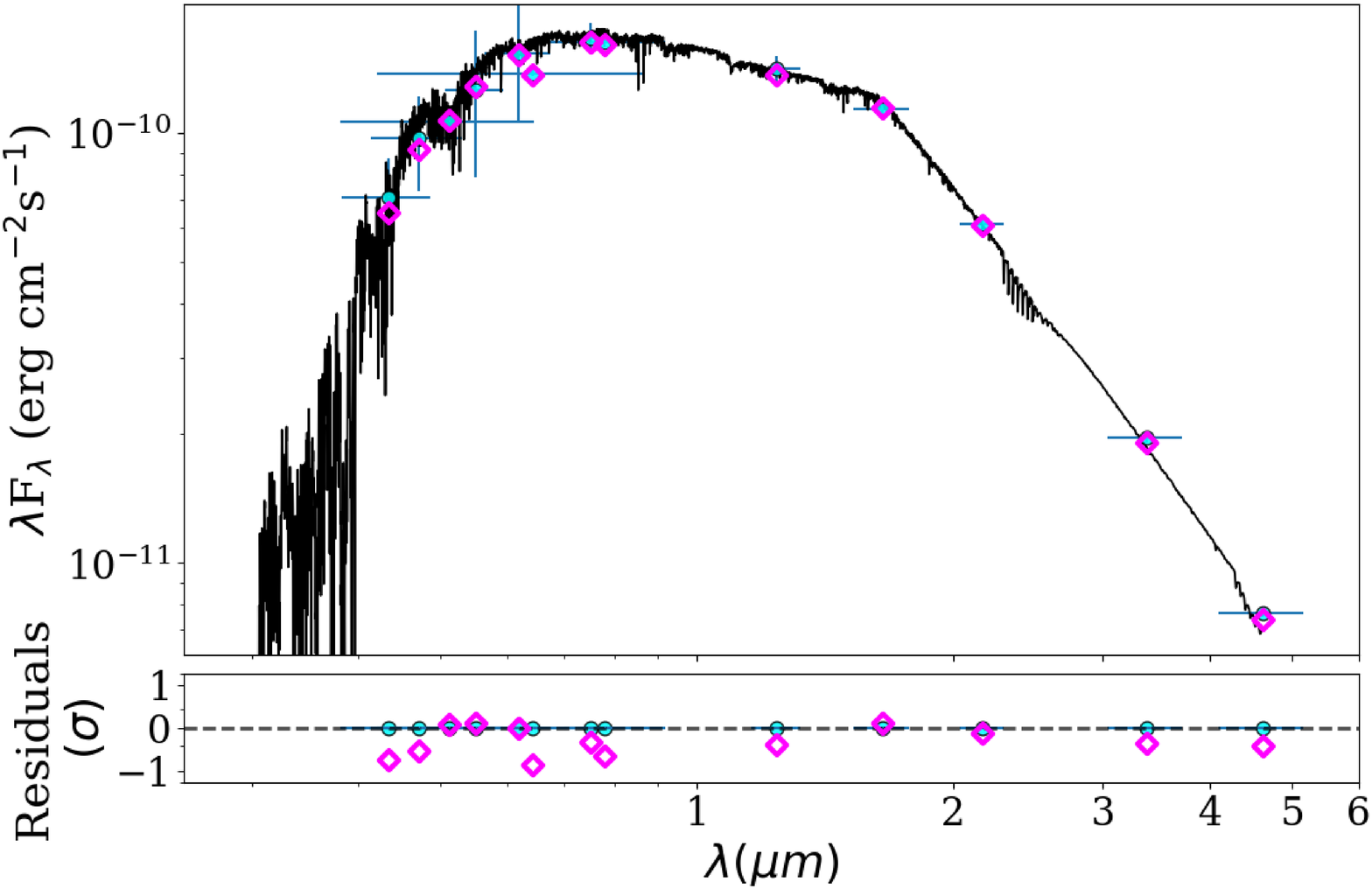}
    \includegraphics[width=0.48\textwidth]{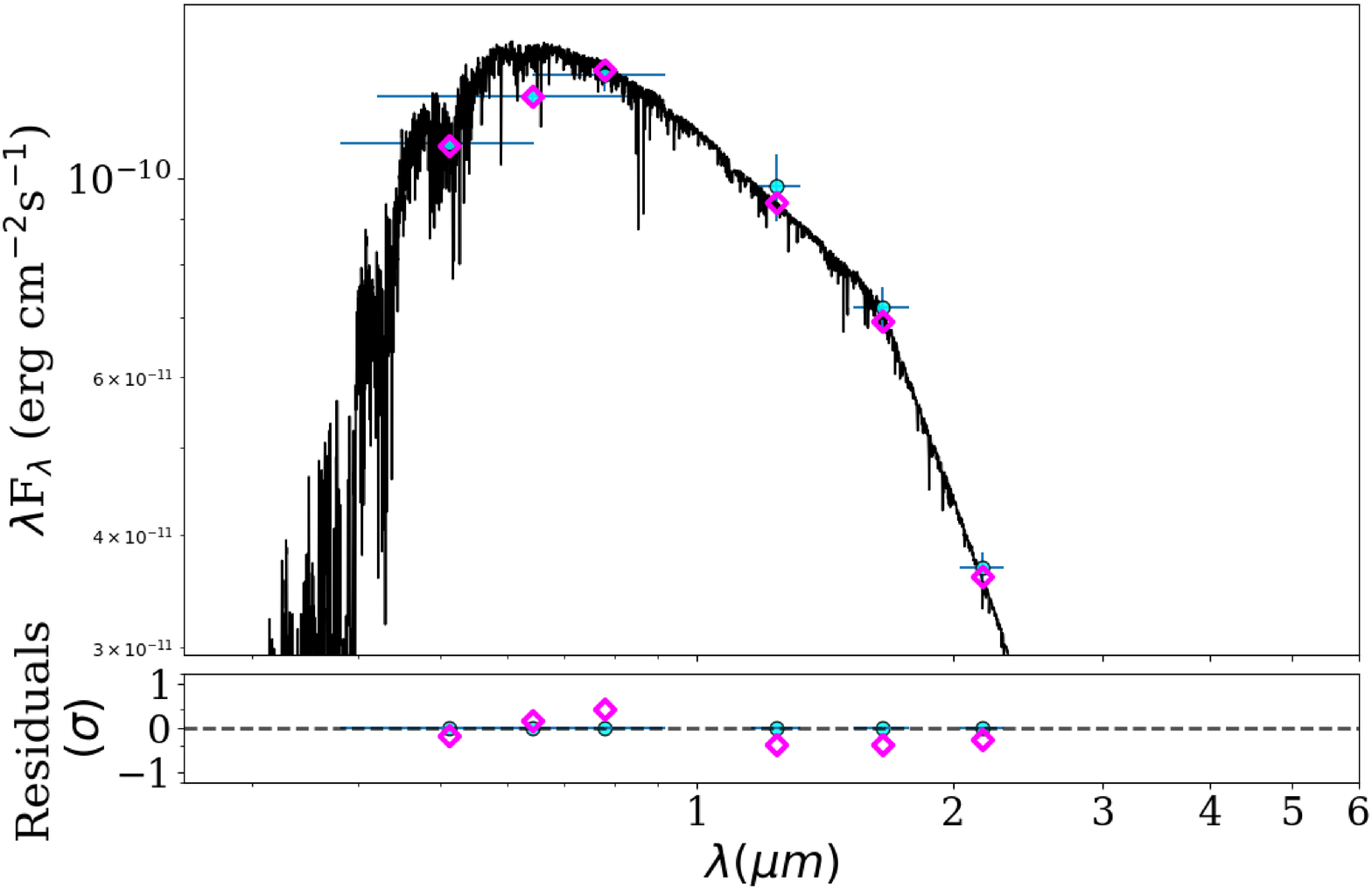}
    \includegraphics[width=0.48\textwidth]{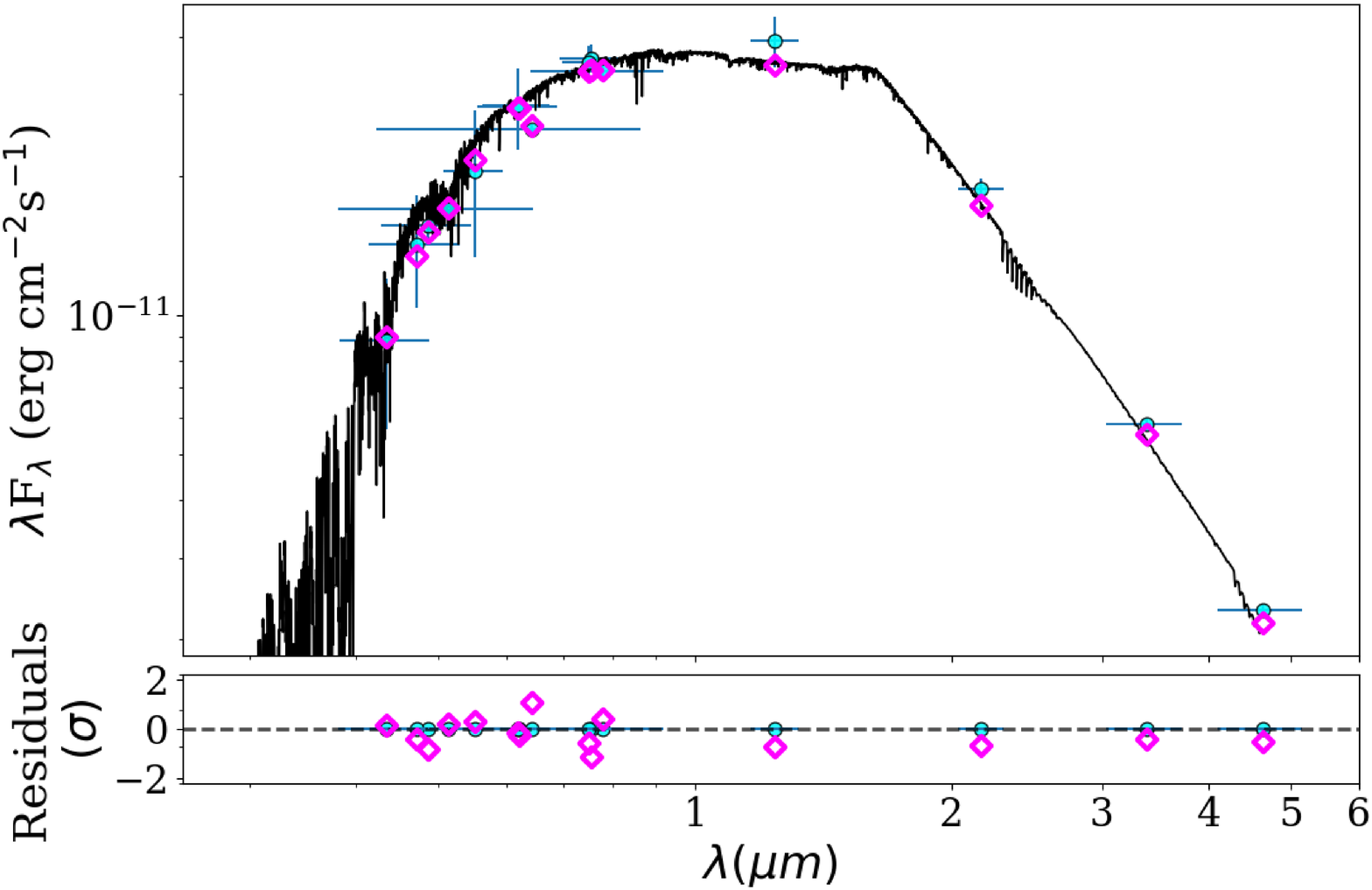}
    \caption{SED fitting of G6084 (top panel), G6081 (middle panel), and G6405 (bottom panel). The photometric data used for fitting (violet circles) are from {\it Gaia} DR2 ($G$, $G_{\rm BP}$, and $G_{\rm RP}$), 2MASS ($J$, $H$, and $K_{\rm S}$), APASS ($B$, $V$, $g$, $r$, and $i$) and WISE ($W$1 and $W$2). The black line represents the best-fit model.}
    \label{8503sed.fig}
\end{figure}

For all the three sources, the best-fit models are from Phoenix.
The gravitational mass of G6084 is about $1.24^{+0.60}_{-0.43}\ M_{\odot}$, while the evolutionary mass is about $0.97^{+0.17}_{-0.05}\ M_{\odot}$.
To G6081, we obtained the gravitational mass of $0.85^{+0.38}_{-0.31}\ M_{\odot}$ and the evolutionary mass of $0.81^{+0.09}_{-0.08},  M_{\odot}$.
For G6405, the gravitational mass is around $0.61^{+0.40}_{-0.25}\ M_{\odot}$ and the evolutionary mass is around $1.02^{+0.14}_{-0.06}\ M_{\odot}$.
Figure \ref{8503sed.fig} shows the SED fitting of the three objects.

\section{Orbital solution}
\label{orbit.sec}

We performed a Keplerian fit using the custom Markov chain Monte Carlo sampler {\it The Joker} \citep{2017ApJ...837...20P}.
{\it The Joker} works well with non-uniform data and allows to identify circular or eccentric orbits.
For G6084 and G6405, only the LAMOST medium-resolution data was used in the fitting; for G6081, both the low- and medium-resolution data were used. 

\begin{figure}
\vspace{-0.6cm}
\center 
\includegraphics[width=0.5\textwidth]{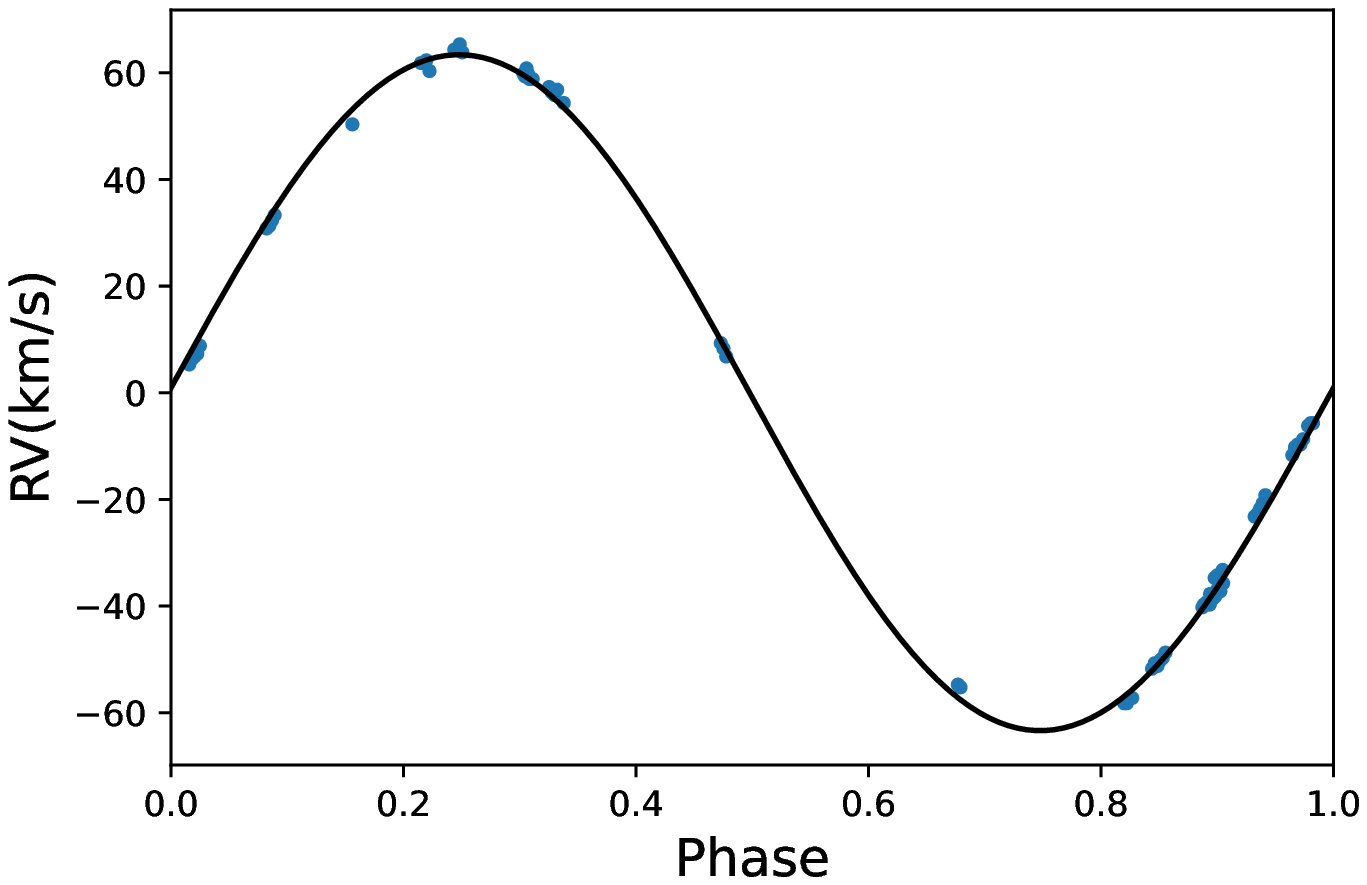}
\includegraphics[width=0.5\textwidth]{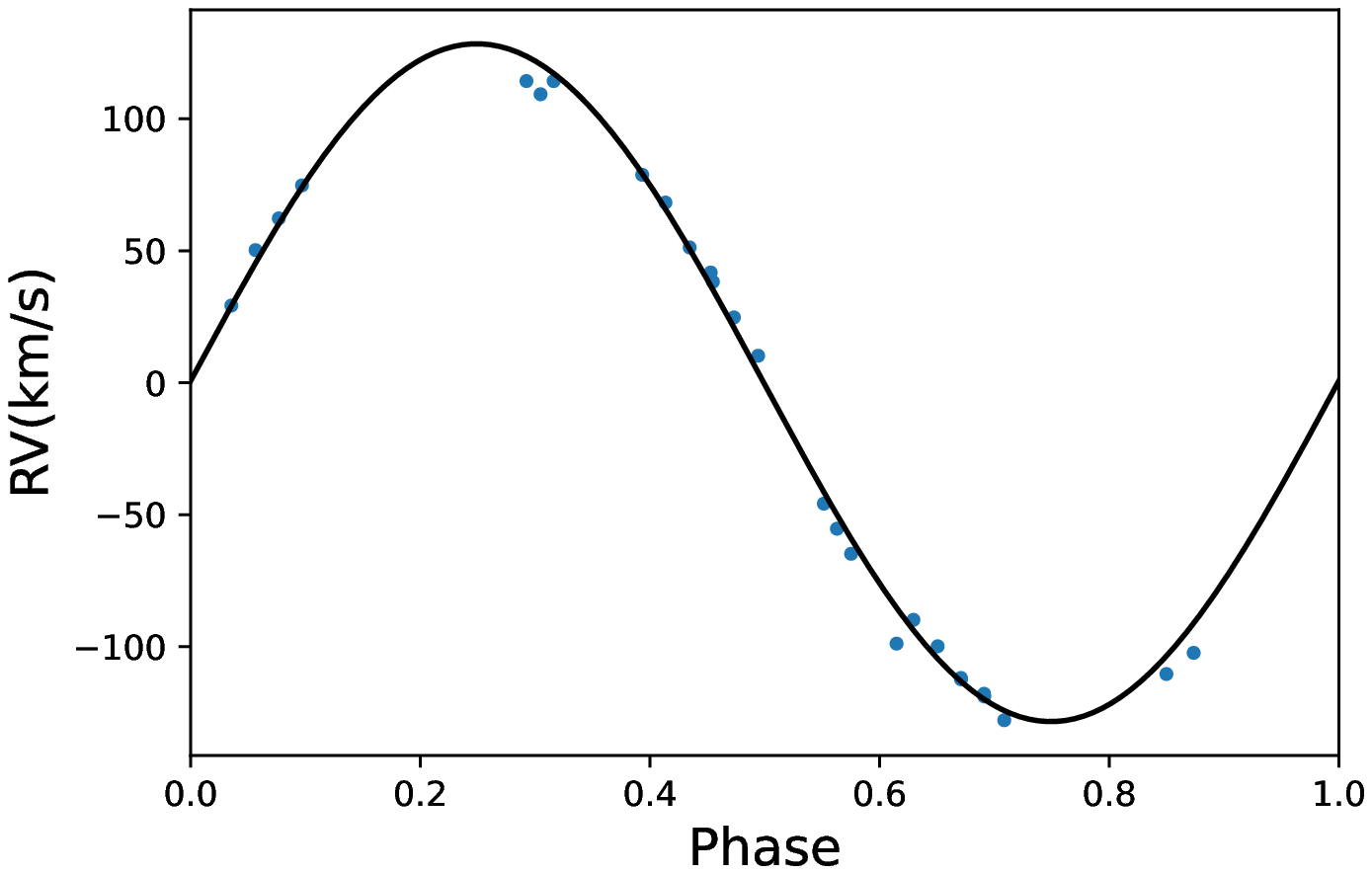}
\includegraphics[width=0.5\textwidth]{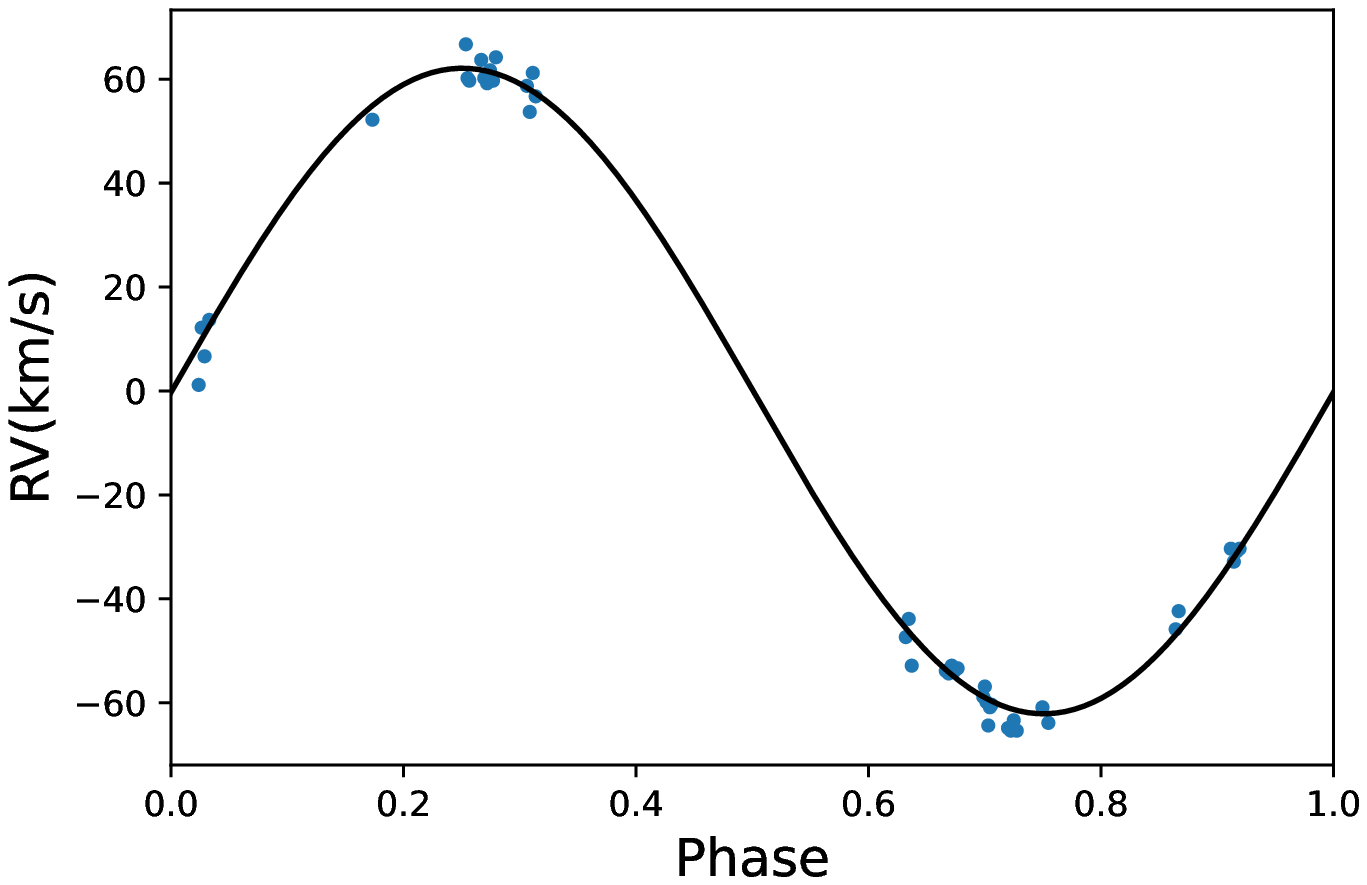}
\caption{Phase-folded RV data (blue dots) and the best-fit RV curves (black lines) of G6084 (top panel), G6081 (middle panel), and G6405 (bottom panel).}
\label{rvb.fig}
\end{figure}

Figure \ref{rvb.fig} shows the RV data and fitted RV curves of the three sources. 
The MCMC results are shown in Figure \ref{8503jokermcmc.fig}, \ref{8050jokermcmc.fig}, and \ref{128jokermcmc.fig} in Appendix \ref{joker_appendix.sec}.
Table \ref{orbital pars.tab} presents all the fit results from {\it The Joker}, including orbital period $P$, eccentricity $e$, argument of the periastron $\omega$, mean anomaly at the first exposure $M_{\rm0}$, semi-amplitude $K$, and systematic RV $\nu_{\rm 0}$.

\begin{table}
\caption{Orbital parameters from {\it the Joker} fitting\label{orbital pars.tab}}
\centering
\setlength{\tabcolsep}{1.pt}
 \begin{tabular}{cccc}
\hline\noalign{\smallskip}
parameters & G6084 & G6081 & G6405 \\
\hline\noalign{\smallskip}
$P$ (day) & $7.02955^{+0.00008}_{-0.00008}$ & $0.789937^{+0.000002}_{-0.000003}$ & $6.42684^{+0.00034}_{-0.00036}$\\
$e$ & $0.011^{+0.002}_{-0.002}$ & $0.016^{+0.005}_{-0.005}$ & $0.03^{+0.01}_{-0.01}$\\
$\omega$ & $1.40^{+0.19}_{-0.20}$ & $3.77^{+0.29}_{-0.30}$ & $2.20^{+0.27}_{-0.23}$\\
$M{\rm 0}$ & $-2.68^{+0.20}_{-0.20}$ & $-2.71^{+0.50}_{-0.24}$ & $-1.90^{+0.26}_{-0.23}$\\
$K_{\rm 1}$ (km/s) & $63.38^{+0.15}_{-0.15}$ & $128.40^{+0.84}_{-0.73}$ & $62.11^{+0.30}_{-0.27}$\\
$\nu{\rm 0}$ (km/s) & $87.51^{+0.12}_{-0.12}$ & $66.08^{+0.37}_{-0.37}$ & $-17.43^{+0.32}_{-0.35}$\\
\noalign{\smallskip}\hline
\end{tabular}
\end{table}

We derived the binary mass function $f$($M$) using the posterior samples from our RV modeling,
\begin{equation}
    f(M) = \frac{M_{2} \, \textrm{sin}^3 i} {(1+q)^{2}} = \frac{P \, K_{1}^{3} \, (1-e^2)^{3/2}}{2\pi G},
\end{equation}
\noindent
where $M_{2}$ is the mass of the invisible star, $q = M_{1}/M_{2}$ is the mass ratio, and $i$ is the system inclination angle. 
By using the mass function, we can calculate the lower limit of the mass of the compact object assuming a maximum inclination angle ($i = 90^{\circ}$).
The mass function of G6084 is $f(M) \approx 0.19$ M$_{\odot}$ and the minimum mass of the unseen object is about 0.94 M$_{\odot}$.
The mass function of G6081 is $f(M) \approx 0.17$ M$_{\odot}$, leading to a minimum mass of the unseen object around 0.69--0.76 M$_{\odot}$.
The mass function of G6405 is $f(M) \approx 0.16$ M$_{\odot}$, while the minimum mass of the invisible object is about 0.69--0.81 M$_{\odot}$.
Table \ref{results.tab} lists all these results in detail.

By using the standard Roche-lobe approximation \citep{1983ApJ...268..368E}
\begin{equation}
\frac{R_1}{a} = 
    \frac{0.49 q^{2/3}}{0.6q^{2/3} + \ln(1 + q^{1/3})},
\label{eq:qi} 
\end{equation}                                 
where $a = (1+q) a_1 = (1+q) P (1-e^2)^{1/2} K_1/(2 \pi {\rm \sin}i)$,
we can calculate a simple solution of the Roche-lobe radius assuming $i=90^{\circ}$.
The calculated Roche-lobe radii are around
7.8 ${\rm R_{\odot}}$ for G6084, 1.6 ${\rm R_{\odot}}$ for G6081, and 6.3--7 ${\rm R_{\odot}}$ for G6405.
For each binary, the Roche-lobe radius of the visible star is larger its physical radius, indicating the visible star hasn't filled its Roche lobe.
Therefore, it is reasonable for us to estimate stellar masses by using single-star evolution models (Section \ref{pmass.sec}).

\begin{table}
\caption{Mass estimations of the binary components \label{results.tab}}
\setlength{\tabcolsep}{4.5pt}
 \begin{tabular}{cccc}
\hline\noalign{\smallskip}
parameters & G6084 & G6081 & G6405 \\
\hline\noalign{\smallskip}
$f$(m) & 0.19 & 0.17 & 0.16\\
\hline\noalign{\smallskip}
\multicolumn{4}{c}{Evolutionary mass of M1 and following calculations}\\
\hline\noalign{\smallskip}
$M_{\rm 1, evo} {\rm (M_{\odot})}$ & $1.17^{+0.10}_{-0.08}$ & $0.83^{+0.01}_{-0.01}$ & $1.00^{+0.08}_{-0.06}$\\
$R_{\rm 1, evo} {\rm (R_{\odot})}$ & $4.60^{+0.20}_{-0.17}$ & $0.967^{+0.005}_{-0.005}$ & $4.86^{+0.10}_{-0.10}$\\
$M_{\rm 2, min} {\rm (M_{\odot})}$ & 0.94 & 0.76 & 0.81\\
$q$ & 1.24 & 1.09 & 1.23\\
\hline\noalign{\smallskip}
\multicolumn{4}{c}{Gravity mass and following calculations}\\
\hline\noalign{\smallskip}
$M_{\rm 1,gra} {\rm (M_{\odot})}$ & $1.18{\pm 0.07}$ & $0.69{\pm 0.11}$ & $0.74{\pm 0.08}$\\
$R_{\rm 1, gra} {\rm (R_{\odot})}$ & $4.50{\pm 0.07}$ & $0.94{\pm 0.02}$ & $4.47{\pm 0.13}$\\
$M_{\rm 2, min} {\rm (M_{\odot})}$ & 0.94 & 0.69 & 0.69\\
$q$ & 1.26 & 1.00 & 1.07\\
\noalign{\smallskip}\hline
\end{tabular}
\smallskip\\
\end{table}

\section{Discussion}
\label{diss.sec}

\subsection{Light curves}
\label{lc.sed}

\begin{figure}
\vspace{-1.8cm}
    \center
    \includegraphics[width=0.5\textwidth]{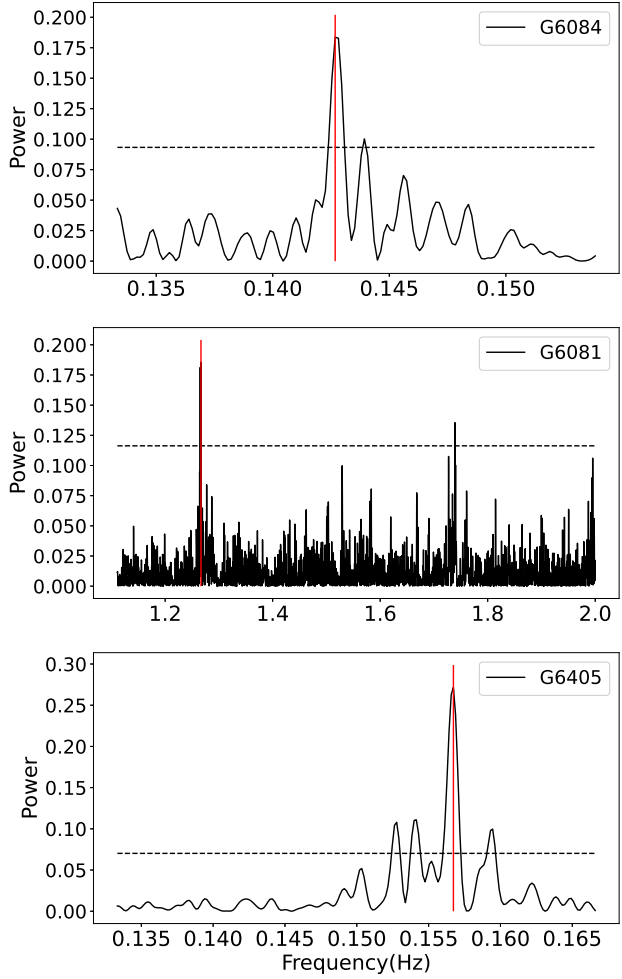}
    \caption{Lomb-Scargle Periodograms of G6084 (top panel), G6081 (middle panel), and G6405 (bottom panel) from ZTF $g$ band observations. Red vertical line is the frequency with the maximum power. Black horizontal dashed line represents the 5$\%$ false alarm probability.}
    \label{ls_plot.fig}
\end{figure}

We collected multi-band LCs from current wide-field photometric TD surveys, and
tried to estimate the orbital periods by using the Lomb-Scargle method \citep{2018ApJS..236...16V}. 
The periods of G6084, G6081, and G6405 were estimated as $\approx$7.00880 day, 0.78951 day, and 6.38246 days, respectively (Figure \ref{ls_plot.fig}), which were similar to those values derived with RV data. 

\begin{figure}
    \center
    \includegraphics[width=0.48\textwidth]{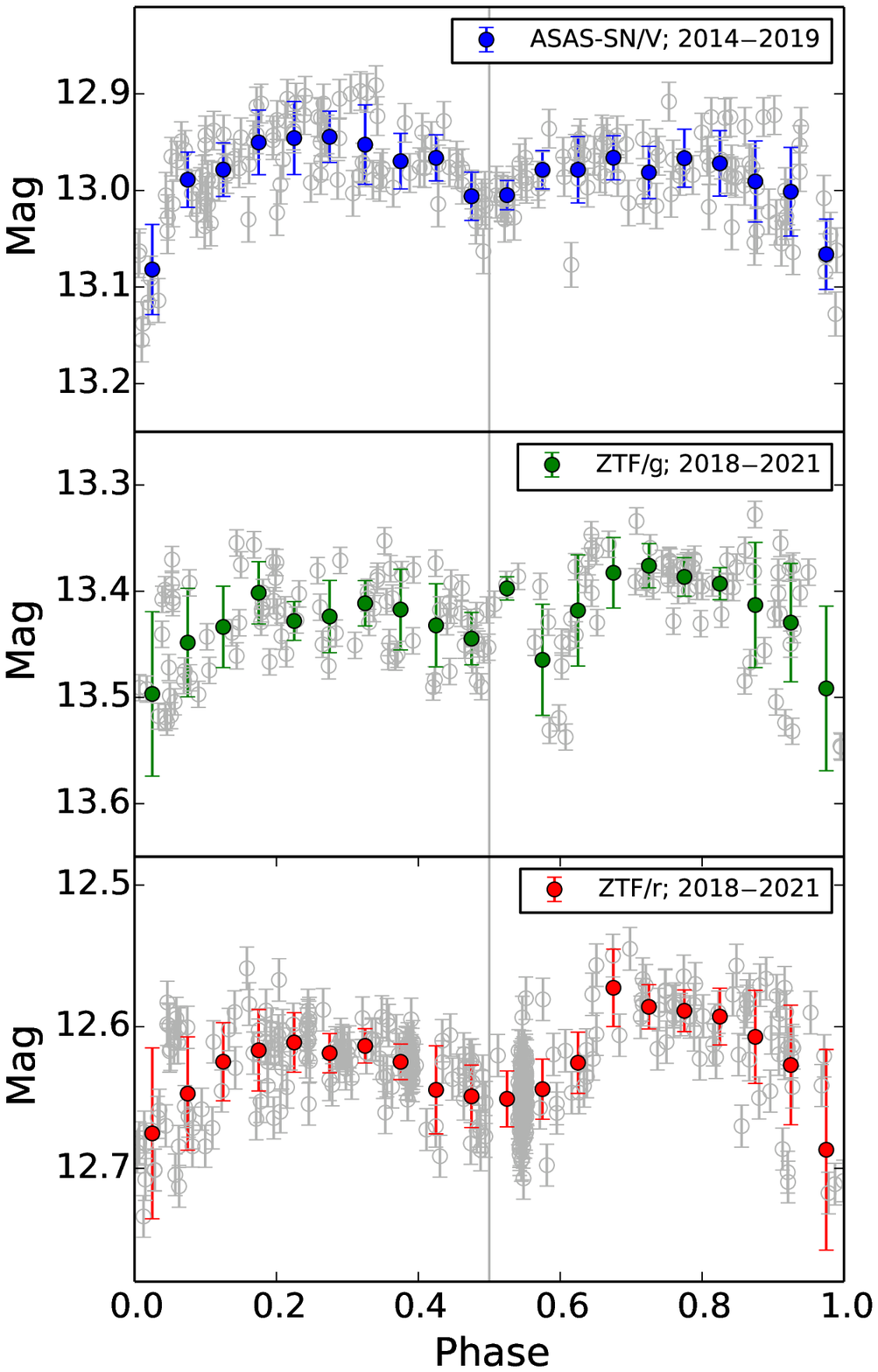}
    \caption{Folded multi-band light curves of G6084.}
    \label{8503lc.fig}
\end{figure}

\begin{figure*}
    \center
    \includegraphics[width=0.98\textwidth]{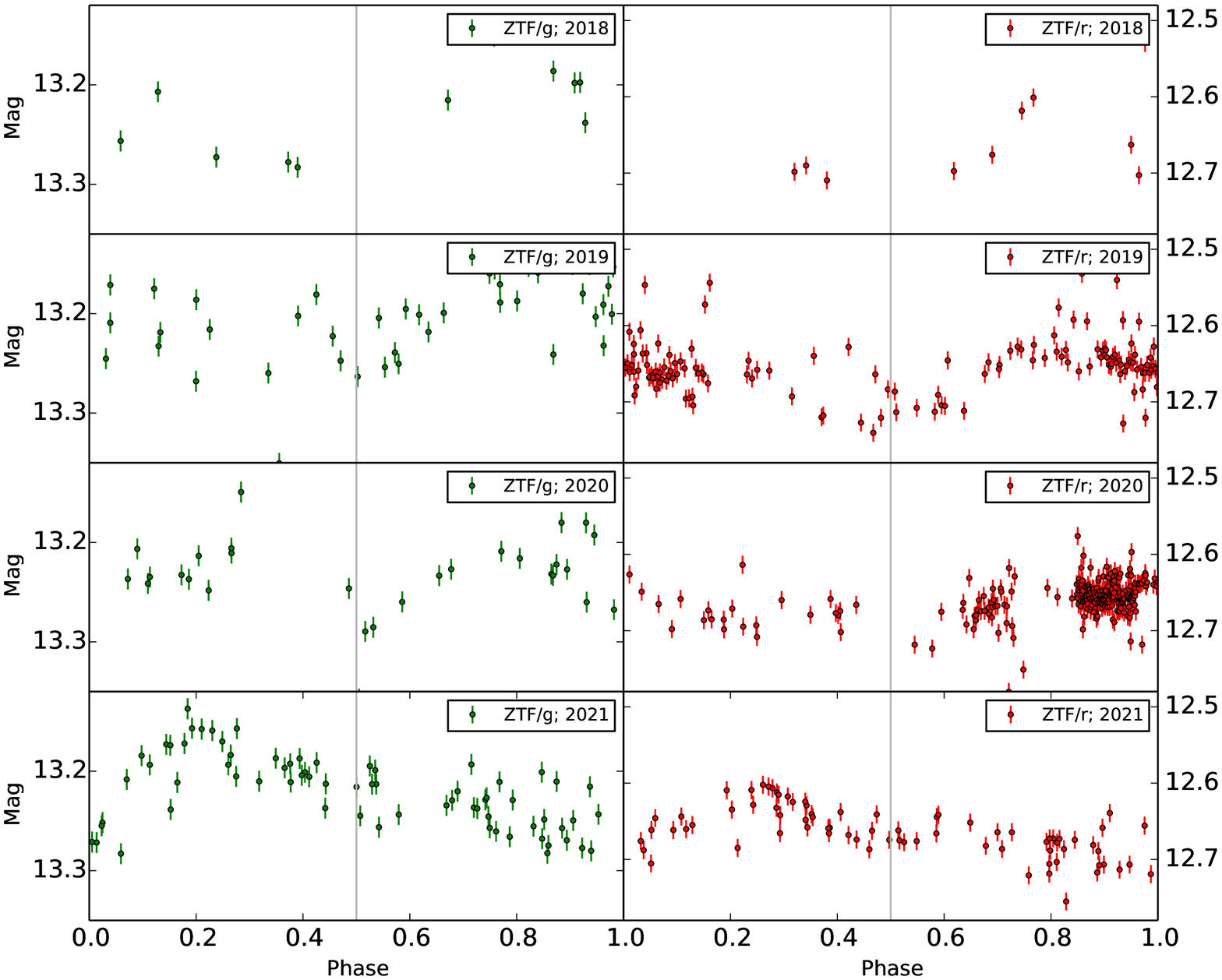}
    \caption{Folded multi-band light curves of G6081.}
    \label{8050lc.fig}
\end{figure*}

\begin{figure*}
    \center
    \includegraphics[width=0.98\textwidth]{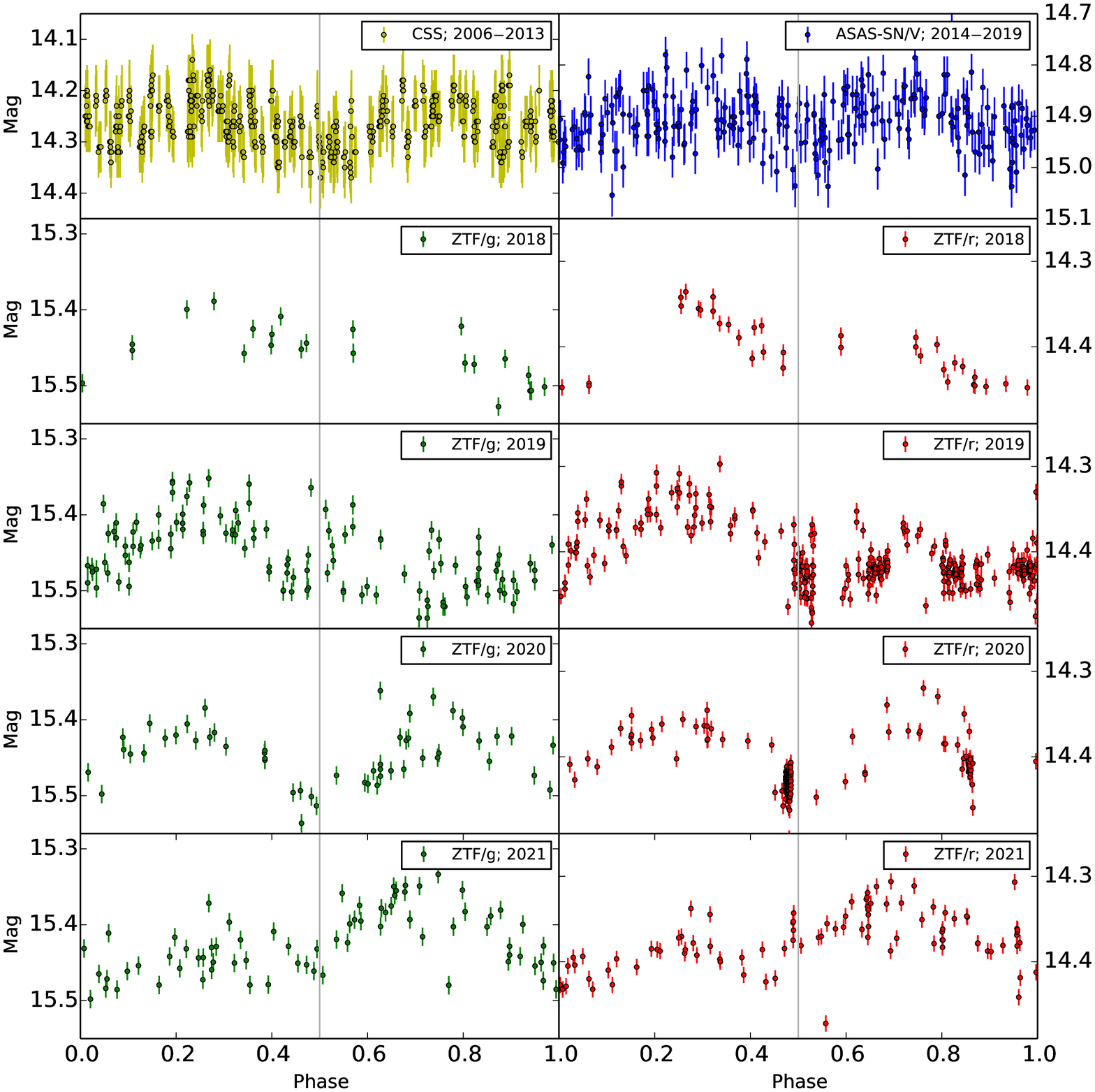}
    \caption{Folded multi-band light curves of G6405.}
    \label{128lc.fig}
\end{figure*}

The LCs (Figure \ref{8503lc.fig}, \ref{8050lc.fig} and \ref{128lc.fig}), folded with the periods estimated from RV data, show typical double-peaked morphology expected for a tidally distorted secondary.
The two asymmetrical peaks in folded LCs suggest an O'Connell effect \citep{2009SASS...28..107W}, 
which may be caused by asymmetric distributed cool starspots \citep{1986ApJ...300..304L}, a hot spot located in an accretion disk \citep{1987AN....308..235A}, or clouds of circumstellar material \citep{2003ChJAA...3..142L}.
For G6081 and G6405, the interchange between the two light maxima in different years are similar to the flip-flop behaviour, which is explained as two active longitudes about 180$^{\rm o}$ apart with alternating levels of spot activity \citep{2001A&A...379L..30K}.
Although the LCs can help constrain the inclination angle, the precision of the data are too low to do the LC fitting.

\subsection{The nature of the unseen object}
\label{nature.sec}

For G6084, the luminosity of the visible star is about ($3.52{\pm 0.09})\times10^{34}$ erg/s.
The minimum luminosity of the unseen object is about ($2.64{\pm 0.06})\times10^{33}$ erg/s following \citep{2018AIPC.2042b0016E},
\begin{equation}
\label{mlr1.eq}
    {\rm log}L = (5.743{\pm 0.413}) \times {\rm log}M - (0.007{\pm 0.026}),
\end{equation}
assuming the unseen object is an main-sequence star of 0.94 M$_{\odot}$. 
The luminosity of the visible star of G6081 is about ($2.42{\pm 0.03})\times10^{33}$ erg/s. Assuming that the unseen object is a dwarf of 0.69 or 0.76 M$_{\odot}$, the luminosity can be calculated as ($5.55{\pm 0.22})\times10^{32}$ erg/s or ($7.79{\pm 0.29})\times10^{32}$ following
\begin{equation}
\label{mlr2.eq}
    {\rm log}L = (4.572{\pm 0.319}) \times {\rm log}M - (0.102{\pm 0.076}).
\end{equation}
For G6405, the luminosity of the visible star is about ($3.40{\pm 0.02})\times10^{34}$ erg/s. The minimum mass of the invisible star (0.69 or 0.81 M$_{\odot}$) leads to a minimum luminosity around ($5.55{\pm 0.22})\times10^{32}$ or ($1.12{\pm 0.02})\times10^{33}$ erg/s following Eq. \ref{mlr2.eq}. In previous studies, G6405 was classified as one RS CVn source due to its spotted feature \citep{2014ApJS..213....9D}.

For all these three sources, the inferred luminosity of the invisible star is about one tenth of the visible star. Although no obvious double-line feature is seen from the LAMOST MRS spectra of the three sources, it's more reasonable to check the binarity of these spectra with the spectral disentangling method. 

\subsubsection{Spectral disentangling}

We used the algorithm of spectral disentangling proposed by \citet{1994A&A...281..286S}. 
For G6081, the maximum mass ratio ($q=M_{\rm 1}/M_{\rm 2}$) is around 1, indicating a secondary with mass close to or larger than that of the primary (which is a main-sequence star). The secondary component can be clearly distinguished from the observed spectra if it's a normal star. Therefore, we only did spectral disentangling for G6084 and G6405.

We first did some test to check the detection limit of the secondary star with this method for our sources.  According to the mass ratio $q$ in Table \ref{results.tab}, we derived approximate effective temperature and surface gravity of the secondary (assuming a main-sequence star) with its minimum mass\footnote{http://www.pas.rochester.edu/\~emamajek/EEM\_dwarf\_UBV-
IJHK\_colors\_Teff.txt}.
We picked out model templates from Phoenix\footnote{https://phoenix.astro.physik.uni-goettingen.de}, reduced their resolution to $R=$ 7500, and applied rotational broadening ($v{\rm sin}i =$ 10 km/s, 50 km/s, 100 km/s and 150 km/s) and absorption. By combining those models with the observational spectra, we got a grid of faked binary spectra for the test.
The spectral disentangling results by visual check are listed in Table \ref{disentangling.tab}. 
When $q$ is 0.8 and $v{\rm sin}i$ $< $ 100 km/s, this method can separate the binary components for both G6804 and G6405. When $q$ is 1.2, for G6804 the spectrum of the secondary can be distinguished but in low significance; 
for G6405 the spectra can't be disentangled, mostly due to the low SNR of the faked spectra.
Considering that the secondary mass in Table \ref{results.tab} is only the minimum value and the mass ratio will decrease as the inclination angle decreases, the disentangling method are still suitable for our targets,
although the constraint on the nature of G6405's secondary may be weak.

\begin{figure*}
    \center
    \includegraphics[width=1\textwidth]{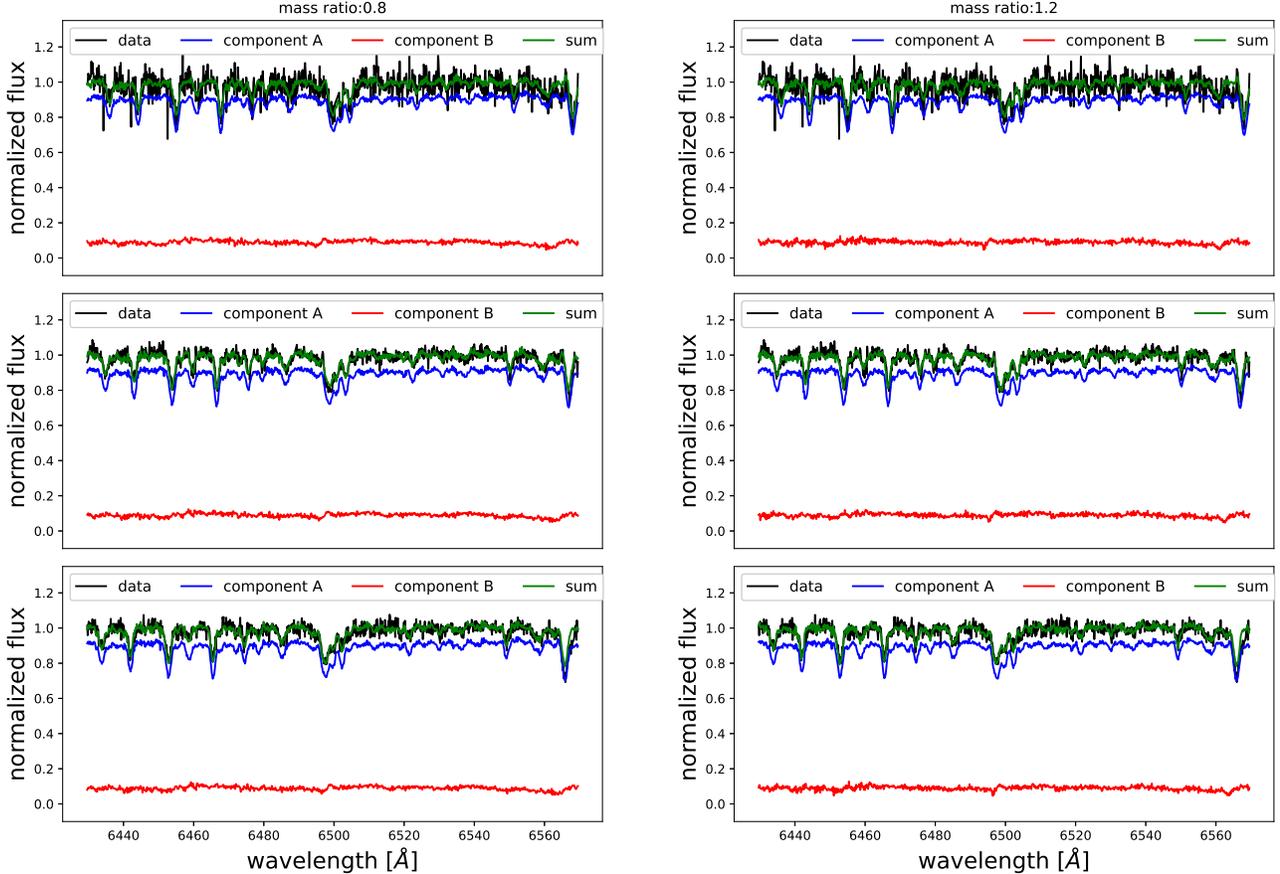}
    \caption{Left panel: spectral disentangling of G6084 with $q =$ 0.8. Right Panel: spectral disentangling of G6084 with $q =$ 1.2. The vertical panels show spectra in different phases (close to the minimum or maximum RV phase of the visible star). The blue lines represent the reconstructed spectra of visible star in those three systems, and the red lines represent the second component in each spectra. The green lines are the sum of the two components, and the black lines mark the observed spectra.}
    \label{disentangling_6084.fig}
\end{figure*}

\begin{figure*}
    \center
    \includegraphics[width=1\textwidth]{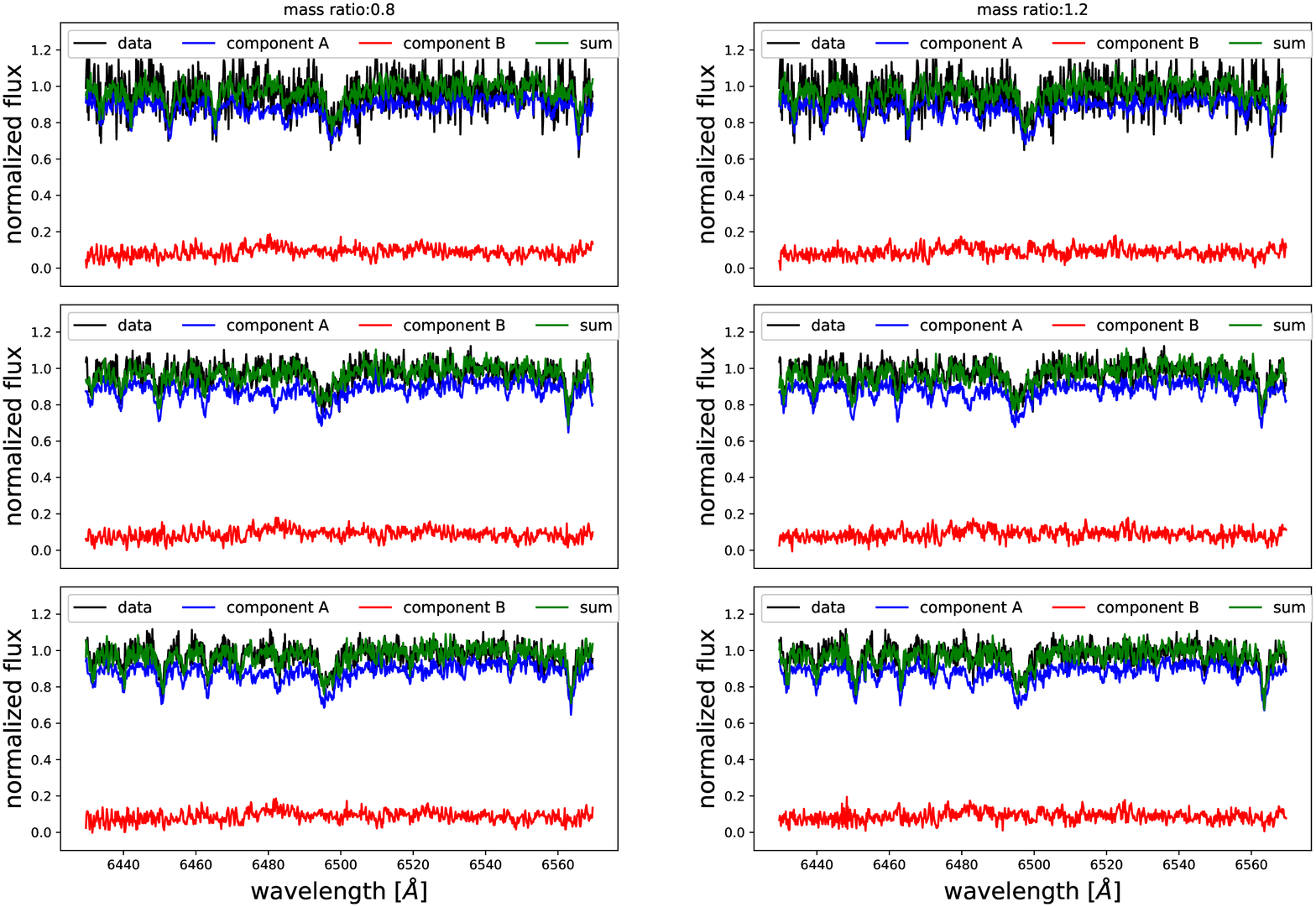}
    \caption{Left panel: spectral disentangling of G6405 with $q =$ 0.8. Right Panel: spectral disentangling of G6405 with $q =$ 1.2. Symbols are as Figure \ref{disentangling_6084.fig}.}
    \label{disentangling_6405.fig}
\end{figure*}

We used the spectrum wavelength ranging from  6400 \AA \ to 6600 \AA \ to disentangle the spectra.
In order to try different sets of RV ratio, we sampled different mass ratio (from 0.8 to 1.2) through MCMC. The disentangling results are shown in Figure \ref{disentangling_6084.fig} and \ref{disentangling_6405.fig}.
No additional component with optical absorption spectrum can be detected, excluding the scenario of normal binary.

\begin{table}
\caption{Spectral disentangling results by visual check. "\ding{52}\ding{52}" means the spectra are well disentangled, "\ding{52}" means the spectra can be disentangled but in low significance, and "\ding{53}" means the spectra can not be disentangled. \label{disentangling.tab}}
\centering
\setlength{\tabcolsep}{4pt}
 \begin{tabular}{cccc}
\hline\noalign{\smallskip}
Object & $q$ & $v{\rm sin}i$ & 
\begin{tabular}{c}Disentangling  \\ results \end{tabular} \\
\hline\noalign{\smallskip}
 \multirow{12}*{G6084} & \multirow{4}*{0.8} & 10 & \ding{52}\ding{52}\\
 & & 50 & \ding{52}\ding{52}\\
 & & 100 & \ding{52}\ding{52}\\
  & & 150 & \ding{53}\\
   \cline{2-4}
&  \multirow{4}*{1.0}  & 10 & \ding{52}\ding{52}\\
& & 50 & \ding{52}\ding{52}\\
& & 100 & \ding{52}\ding{52}\\
 & & 150 & \ding{53}\\
  \cline{2-4}
&  \multirow{4}*{1.2}  & 10 & \ding{52}\\
& & 50 & \ding{52}\\
& & 100 & \ding{52}\\
 & & 150 & \ding{53}\\
\hline
 \multirow{12}*{G6405} & \multirow{4}*{0.8}  & 10 & \ding{52}\ding{52}\\
 & & 50 & \ding{52}\ding{52}\\
 & & 100 & \ding{52}\ding{52}\\
  & & 150 & \ding{53}\\
  \cline{2-4}
  & \multirow{4}*{1.0} & 10 & \ding{52}\\
 & & 50 & \ding{52}\\
 & & 100 & \ding{53}\\
 & & 150 & \ding{53}\\
  \cline{2-4}
 & \multirow{4}*{1.2} & 10 & \ding{53}\\
 & & 50 & \ding{53}\\
 & & 100 & \ding{53}\\
 & & 150 & \ding{53}\\

\noalign{\smallskip}\hline
\end{tabular}
\end{table}

\subsubsection{X-ray upper limit}
\label{star.uvxray}

The sky fields of the three sources have been covered by ROSAT during its X-ray sky survey, although without any detection of them.
The observations were performed with the instrument Position Sensitive Proportional Counter (hereafter PSPC).
Here we tried to give upper limit estimations of these sources by using the archival PSPC data.

For G6081 and G6405, we used the observations ObsID rp700887n00 and rs931310n00, respectively.
The radius of the photometric aperture was set as 30$^{\prime\prime}$ and an annulus around it was used to estimate the background emission. 
For G6081, we obtained an upper limit of $\approx$0.00024 count s$^{-1}$ in the 0.2--8 keV range.
The hydrogen column density $N_{\rm H}$ can be calculated following the standard linear relation between $N_{\rm H}$ and the reddening $N_{\rm H} = $ 5.8$\times10^{21}$ cm$^{-2}$ $E(B-V)$.
By using the WebPIMMS\footnote{https://heasarc.gsfc.nasa.gov/cgi-bin/Tools/w3pimms/w3pimms.pl}, we derived an unabsorbed flux of 5.7$\times 10^{-15}\ \rm erg\, s^{-1}\, cm^{-2}$ in the 0.2--8 keV range assuming $\tau =$1.5 for a power-law spectrum, or 2.9$\times 10^{-15}\ \rm erg\, s^{-1}\, cm^{-2}$  assuming $\tau =$2.
At a distance $d=0.32\,$ kpc (Section \ref{para.sec}), the flux corresponds to a luminosity of $L_X$ $\lesssim$ 7.0$\times$10$^{28}$ $\rm erg\, s^{-1}$ or
3.5$\times$10$^{28}$ $\rm erg\, s^{-1}$.
For G6405, the upper limit count rate is about 0.000052 count s$^{-1}$.
The estimated unabsorbed flux is around 2.4$\times 10^{-15}\ \rm erg\, s^{-1}\, cm^{-2}$ ($\tau =$1.5) or 1.8$\times 10^{-15}\ \rm erg\, s^{-1}\, cm^{-2}$ ($\tau =$2),
corresponding to a luminosity of $L_X$ $\lesssim$ 1.1$\times$10$^{30}$ $\rm erg\, s^{-1}$ or 8.5$\times$10$^{29}$ $\rm erg\, s^{-1}$.

For G6084, we used the observation Obsid rs931310n00 to determine the upper limit count rate.
Due to the quite few photons, a larger radius (60$^{\prime\prime}$) was used to do the aperture photometry.
A count rate of 0.000054 count s$^{-1}$ leads to an unabsorbed flux of 1.3$\times 10^{-15} \rm erg\, s^{-1}\, cm^{-2}$ ($\tau =$1.5) or 6.5$\times 10^{-16} \rm erg\, s^{-1}\, cm^{-2}$ ($\tau =$2).
This corresponds to a luminosity of $L_X$ $\lesssim$ 1.9$\times$10$^{29}$ $\rm erg\, s^{-1}$ or 9.4$\times$10$^{28}$ $\rm erg\, s^{-1}$.

\subsubsection{UV emission}

Both G6084 and G6081 have been observed by the {\it Galaxy Evolution Explorer} (hereafter {\it GALEX}).
For G6084, the {\it GALEX} catalog presents a NUV magnitude of 19.22$\pm$0.22 mag.
The FUV magnitude of G6081 is 22.51$\pm$0.18, however, the {\it GALEX} catalog doesn't present a NUV magnitude due to the contamination of one nearby bright star.

We tried to do the aperture photometry using the Python module Photutils.
The magnitudes were estimated with the formulae\footnote{https://asd.gsfc.nasa.gov/archive/galex/FAQ/counts\_back-\\ground.html}:
\begin{equation}
    m_{\rm FUV, AB} = -2.5 \times {\rm log10(Count\ Rate)} + 18.82
\end{equation}
   and
\begin{equation}
    m_{\rm NUV, AB} = -2.5 \times {\rm log10(Count\ Rate)} + 20.02.
\end{equation}
The NUV magnitude of G6084 is 19.46$\pm$0.02 mag; an upper limit of the FUV band was derived as 23.64$\pm$1.14 mag, respectively.
The FUV and NUV magnitudes of G6081 are 22.55$\pm$0.26 and 18.35$\pm$0.03 mag. 

In addition, ROSAT consists one Wide Field Camera (WFC) covering the wavelength ranges of 60--140 \AA\ and 110--200 \AA. 
We investigated the ROSAT/WFC images for G6405 but no UV signal was found.

\section{Summary}
\label{sum.sec}

We presented the discovery of three binaries with possible compact components (i.e., G6084, G6081 and G6405) by using the LAMOST LRS and MRS data from our spectroscopic TD survey.
G6084 is a binary with an orbital period of $P =$ 7.03 day containing a K-type subgiant. This visible primary has a mass of $\approx$ 1.18$\pm0.07$ M$_{\odot}$. By fitting the RV curve, we calculated the mass function of 0.19 M$_{\odot}$ and the minimum mass of the unseen secondary of 0.94 M$_{\odot}$.
G6081 has an orbital period of $P =$ 0.79 day. The visible star is a G-type dwarf with a mass of $\approx$0.83$\pm0.01$ (or 0.69$\pm0.11$) M$_{\odot}$. The binary mass function is 0.17 M$_{\odot}$, and the minimum mass of the unseen object is estimated as $\approx$0.76 (or 0.69) M$_{\odot}$. 
G6405 is a binary with orbital period $P =$ 6.43 day. The visible star is a K-type subgiant with a mass of $\approx$1.00$\pm0.07$ (or 0.74$\pm0.08$) M$_{\odot}$. The mass function is 0.16 M$_{\odot}$. The compact object has a minimum mass of $\approx$0.81 (or 0.69) M$_{\odot}$. 
None of the visible stars has filled its Roche lobe.

The LCs and RVs show the same periods. However, due to the poor quality of the photometric data, we did not constrain the inclination angle of these binary systems.
No double-line feature can be seen in the LAMOST medium-resolution spectra.
Furthermore, we did spectral disentangling for our sources and found no additional component with absorption spectra, supporting the scenario that these sources are binary systems with compact components (i.e., including a white dwarf or neutron star).
Note that the nature of the secondary of G6405 is weakly constrained by the spectral disentangling method, due to the low SNR of the observed spectra.
The ROSAT data show no X-ray detection of these sources. The upper limits of X-ray luminosity of them range from 10$^{28}$ $\rm erg\, s^{-1}$ to 10$^{30}$ $\rm erg\, s^{-1}$.
Both G6048 and G6405 have bright UV counterparts, suggesting that they are more likely WDs rather than NSs.

Besides the four $K$2 plates, we are also carrying out a TD survey of another four $K$2 plates from 2020. More binaries containing compact objects can be found based on the LAMOST TD survey, which will help us in better understanding the late evolution of massive stars (in binaries).

\section*{acknowledgements}
Guoshoujing Telescope (the Large Sky Area Multi-Object Fiber Spectroscopic Telescope LAMOST) is a National Major Scientific Project built by the Chinese Academy of Sciences. Funding for the project has been provided by the National Development and Reform Commission. LAMOST is operated and managed by the National Astronomical Observatories, Chinese Academy of Sciences.
This work presents results from the European Space Agency (ESA) space mission Gaia. Gaia data are being processed by the Gaia Data Processing and Analysis Consortium (DPAC). Funding for the DPAC is provided by national institutions, in particular the institutions participating in the Gaia MultiLateral Agreement (MLA). The Gaia mission website is https://www.cosmos.esa.int/gaia. The Gaia archive website is https://archives.esac.esa.int/gaia.
We acknowledge use of the VizieR catalogue access tool, operated at CDS, Strasbourg, France, and of Astropy, a community-developed core Python package for Astronomy (Astropy Collaboration, 2013).
This research made use of Photutils \citet{2020zndo...4044744B}, an Astropy package for detection and photometry of astronomical sources.
This work was supported by National Science Foundation of China (NSFC) under grant numbers 11988101/11933004, National Key Research and Development Program of China (NKRDPC) under grant numbers 2019YFA0405504 and 2019YFA0405000, and Strategic Priority Program of the Chinese Academy of Sciences under grant number XDB41000000.
S. W. and H.-L. Y. and  acknowledges support from the Youth Innovation Promotion Association of the CAS (id. 2019057 and 2020060, respectively).


\bibliographystyle{aasjournal}
\bibliography{main.bbl}

\clearpage
\appendix
\renewcommand*\thetable{\Alph{section}.\arabic{table}}
\renewcommand*\thefigure{\Alph{section}\arabic{figure}}

\section{LAMOST RV measurements for our targets.}

Here we presented the LAMOST RV values used in this paper.

\setcounter{table}{0}
 \begin{table}[h]
 \caption{Barycentric-corrected RV values of G6084 from LAMOST observations. \label{lamost1.tab}}
  \renewcommand{\arraystretch}{0.85}
 \setlength{\tabcolsep}{4.5pt}
 \begin{tabular}{ccccc|ccccc}
 \hline\noalign{\smallskip}
 BMJD & RV & Uncertainty & {\it SNR} & Resolution & BMJD & RV & Uncertainty & {\it SNR} & Resolution\\
 (day) & (km/s) & (km/s) &  &  & (day) & (km/s) & (km/s) & & \\
 \hline\noalign{\smallskip}
 58798.83429 & 52.78 & 0.7 & 21.12 & MRS &	58889.71963 & 30.27 & 0.65 & 19.14 & MRS \\
58798.85096 & 53.28 & 0.6 & 16.7 & MRS &	58895.69519 & 32.77 & 0.65 & 19.51 & MRS \\
58798.86694 & 52.78 & 0.75 & 10.01 & MRS &	58895.71117 & 32.27 & 0.65 & 19.53 & MRS \\
58798.88291 & 54.28 & 0.8 & 8.2 & MRS &     58912.60353 & 118.32 & 0.55 & 40.35 & MRS \\
58805.83359 & 47.78 & 0.65 & 17.98 & MRS &	58912.62019 & 118.82 & 0.55 & 35.22 & MRS \\
58805.84956 & 48.78 & 0.6 & 23.16 & MRS &	58912.63617 & 119.82 & 0.6 & 33.16 & MRS \\
58805.86623 & 49.28 & 0.65 & 21.79 & MRS &	58912.65214 & 120.83 & 0.6 & 28.52 & MRS \\
58805.88221 & 50.78 & 0.6 & 20.14 & MRS &	58920.56694 & 149.34 & 0.65 & 23.29 & MRS \\
58805.89818 & 50.28 & 0.7 & 18.67 & MRS &	58920.58292 & 149.34 & 0.7 & 16.45 & MRS \\
58805.91485 & 51.78 & 0.6 & 17.07 & MRS &	58920.59889 & 149.84 & 0.8 & 10.51 & MRS \\
58819.84877 & 47.28 & 0.6 & 24.04 & MRS &	58920.61833 & 147.84 & 1.05 & 6.64 & MRS \\
58819.85780 & 47.78 & 0.6 & 26.67 & MRS &	58950.49786 & 96.81 & 0.55 & 28.81 & MRS \\
58819.86752 & 47.78 & 0.6 & 26.04 & MRS &	58950.51383 & 95.81 & 0.55 & 24.26 & MRS \\
58819.87655 & 48.28 & 0.6 & 23.53 & MRS &	58950.53050 & 94.31 & 0.6 & 22.73 & MRS \\
58819.88627 & 48.28 & 0.6 & 21.42 & MRS &	59150.87673 & 81.3 & 0.55 & 23.03 & MRS \\
58819.89530 & 49.78 & 0.6 & 21.61 & MRS &	59150.89271 & 81.8 & 0.6 & 26.57 & MRS \\
58819.90433 & 49.78 & 0.6 & 23.9 & MRS &	59150.90649 & 81.8 & 0.7 & 7.85 & MRS \\
58829.80167 & 147.34 & 0.65 & 15.89 & MRS &	59180.86159 & 151.84 & 0.65 & 20.32 & MRS \\
58829.81140 & 146.84 & 0.6 & 17.62 & MRS &	59180.87826 & 151.84 & 0.6 & 26.94 & MRS \\
58829.82043 & 148.34 & 0.65 & 14.62 & MRS &	59180.89424 & 152.85 & 0.65 & 17.17 & MRS \\
58829.82945 & 147.34 & 0.65 & 16.6 & MRS &	59180.91021 & 151.34 & 0.7 & 14.02 & MRS \\
58829.83918 & 146.34 & 0.65 & 15.73 & MRS &	59206.79000 & 64.29 & 0.65 & 25.64 & MRS \\
58829.84821 & 146.34 & 0.7 & 16.19 & MRS &	59206.80598 & 64.79 & 0.65 & 29.24 & MRS \\
58829.85793 & 146.34 & 0.65 & 15.97 & MRS &	59206.82264 & 65.79 & 0.65 & 28.21 & MRS \\
58861.72151 & 35.77 & 0.65 & 19.16 & MRS &	59206.83862 & 66.79 & 0.6 & 30.15 & MRS \\
58861.73749 & 36.77 & 0.65 & 18.87 & MRS &	59206.85459 & 68.29 & 0.65 & 29.84 & MRS \\
58861.75415 & 36.27 & 0.65 & 20.58 & MRS &	59237.67106 & 144.84 & 0.6 & 18.25 & MRS \\
58861.77013 & 37.27 & 0.65 & 17.66 & MRS &	59237.68773 & 143.84 & 0.65 & 17.6 & MRS \\
58861.78610 & 37.77 & 0.65 & 19.11 & MRS &	59237.70370 & 143.34 & 0.75 & 8.61 & MRS \\
58861.80277 & 38.77 & 0.65 & 18.6 & MRS &	59237.71967 & 144.34 & 0.6 & 18.4 & MRS \\
58883.65859 & 75.8 & 0.6 & 29.5 & MRS &	59237.75995 & 141.84 & 0.65 & 16.48 & MRS \\
58883.67457 & 77.3 & 0.55 & 28.01 & MRS &	59263.61317 & 92.81 & 0.6 & 13.64 & MRS \\
58883.69123 & 77.8 & 0.55 & 26.38 & MRS &	59263.62915 & 93.81 & 0.6 & 15.04 & MRS \\
58883.70720 & 77.8 & 0.55 & 24.04 & MRS &	59263.64512 & 94.31 & 0.6 & 18.36 & MRS \\
58883.72387 & 78.8 & 0.55 & 25.5 & MRS &	59263.66178 & 94.81 & 0.55 & 22.23 & MRS \\
58889.67102 & 29.27 & 0.7 & 17.43 & MRS &	59263.67775 & 96.31 & 0.55 & 24.79 & MRS \\
58889.68699 & 29.27 & 0.7 & 15.13 & MRS &	59271.62953 & 137.84 & 0.75 & 10.52 & MRS \\
58889.70297 & 30.27 & 0.7 & 17.71 & MRS \\	
 \noalign{\smallskip}\hline
  \end{tabular}
  \end{table}
  
  \clearpage
 
 \begin{table}
 \caption{Barycentric-corrected RV values of G6081 from LAMOST observations. \label{lamost2.tab}}
 \setlength{\tabcolsep}{4.5pt}
\begin{tabular}{ccccc|ccccc}
 \hline\noalign{\smallskip}
 BMJD & RV & Uncertainty & {\it SNR} & Resolution & BMJD & RV & Uncertainty & {\it SNR} & Resolution\\
 (day) & (km/s) & (km/s) &  &  & (day) & (km/s) & (km/s) & & \\
 \hline\noalign{\smallskip}
58837.76259 & 180.36 & 2.95 & 14.59 & LRS &	58912.62026 & 116.32 & 1.4 & 8.75 & MRS \\
58837.77231 & 175.36 & 2.9 & 14.75 & LRS &	58912.63624 & 128.33 & 1.35 & 10.11 & MRS \\
58837.78134 & 180.36 & 2.75 & 15.29 & LRS &	58912.65221 & 140.84 & 1.7 & 7.5 & MRS \\
58843.73242 & -44.28 & 2.6 & 17.26 & LRS &	59150.87661 & -46.28 & 1.1 & 31.64 & MRS \\
58843.75117 & -36.27 & 2.35 & 19.18 & LRS &	59150.89258 & -52.78 & 0.95 & 36.94 & MRS \\
59191.85897 & 20.26 & 1.55 & 58.56 & LRS &	59150.90636 & -61.79 & 1.45 & 12.31 & MRS \\
59191.86799 & 10.76 & 1.5 & 61.26 & LRS &	59180.86148 & -23.76 & 1.05 & 25.34 & MRS \\
59191.87772 & 1.25 & 1.45 & 65.06 & LRS &	59180.87815 & -33.77 & 0.95 & 32.24 & MRS \\
58819.84866 & -32.77 & 8.25 & 3.27 & MRS &	59180.89412 & -45.78 & 1.05 & 21.31 & MRS \\
58883.65860 & 144.84 & 1.15 & 8.86 & MRS &	59180.91010 & -51.78 & 1.15 & 17.18 & MRS \\
58883.67457 & 134.33 & 1.1 & 11.94 & MRS &	59206.78993 & 107.82 & 0.95 & 32.79 & MRS \\
58883.69124 & 117.32 & 1.1 & 12.07 & MRS &	59206.80590 & 90.81 & 1.0 & 36.48 & MRS \\
58883.70721 & 104.32 & 1.2 & 9.35 & MRS &	59206.82257 & 76.3 & 0.95 & 37.39 & MRS \\
58912.60360 & 95.31 & 1.05 & 19.74 & MRS \\
 \noalign{\smallskip}\hline
 \end{tabular}
 \end{table}
 
 \clearpage

 \begin{table}
 \caption{Barycentric-corrected RV values of G6405 from LAMOST observations. \label{lamost3.tab}}
 \setlength{\tabcolsep}{4.5pt}
  \begin{tabular}{ccccc|ccccc}
 \hline\noalign{\smallskip}
 BMJD & RV & Uncertainty & {\it SNR} & Resolution & BMJD & RV & Uncertainty & {\it SNR} & Resolution\\
 (day) & (km/s) & (km/s) &  &  & (day) & (km/s) & (km/s) & & \\
 \hline\noalign{\smallskip}
58801.71828 & -47.78 & 1.05 & 5.47 & MRS &	58890.46206 & -82.3 & 0.95 & 9.75 & MRS \\
58801.73495 & -50.28 & 1.05 & 4.97 & MRS &	58890.47803 & -82.8 & 0.95 & 9.97 & MRS \\
58801.75092 & -48.28 & 1.1 & 3.96 & MRS &	58890.49400 & -80.8 & 0.9 & 9.03 & MRS \\
58801.76674 & -47.78 & 1.65 & 3.31 & MRS &	58890.51067 & -82.8 & 0.95 & 7.99 & MRS \\
58819.63005 & -76.3 & 1.25 & 6.12 & MRS &	59123.78081 & -16.26 & 1.15 & 4.29 & MRS \\
58819.63908 & -74.3 & 1.15 & 6.4 & MRS &	59123.79748 & -5.25 & 1.5 & 4.66 & MRS \\
58819.64810 & -77.3 & 1.15 & 5.35 & MRS &	59123.81345 & -10.76 & 1.7 & 3.94 & MRS \\
58819.65783 & -81.8 & 1.2 & 5.22 & MRS &	59123.83915 & -3.75 & 2.45 & 3.06 & MRS \\
58819.66685 & -78.3 & 1.2 & 5.68 & MRS &	59147.72603 & -78.3 & 1.6 & 3.97 & MRS \\
58819.67588 & -77.8 & 1.25 & 5.61 & MRS &	59147.75867 & -81.3 & 1.85 & 3.27 & MRS \\
58829.62353 & 49.28 & 1.65 & 3.04 & MRS &	59180.59676 & -63.29 & 1.35 & 3.49 & MRS \\
58829.63256 & 42.78 & 1.5 & 3.33 & MRS &	59180.61343 & -59.79 & 1.6 & 3.22 & MRS \\
58829.64228 & 42.28 & 1.5 & 3.37 & MRS &	59189.61118 & 46.28 & 1.1 & 7.07 & MRS \\
58835.53361 & 34.77 & 1.55 & 3.82 & MRS &	59189.62785 & 42.78 & 2.05 & 5.01 & MRS \\
58851.55712 & -71.29 & 0.85 & 11.56 & MRS &	59189.64382 & 41.78 & 1.45 & 5.45 & MRS \\
58851.57309 & -71.79 & 0.85 & 11.66 & MRS &	59189.65979 & 44.28 & 1.4 & 4.74 & MRS \\
58851.58975 & -70.29 & 0.85 & 10.46 & MRS &	59189.67646 & 42.28 & 1.45 & 4.3 & MRS \\
58851.60572 & -71.29 & 0.95 & 11.26 & MRS &	59189.69243 & 46.78 & 1.4 & 3.99 & MRS \\
58851.62239 & -70.79 & 0.95 & 10.45 & MRS &	59215.57107 & 41.28 & 1.1 & 7.7 & MRS \\
58883.47039 & -64.79 & 1.45 & 3.26 & MRS &	59215.58704 & 36.27 & 1.05 & 7.16 & MRS \\
58883.48636 & -61.29 & 1.2 & 3.41 & MRS &	59215.60371 & 43.78 & 1.15 & 6.31 & MRS \\
58883.50302 & -70.29 & 1.3 & 3.39 & MRS &	59215.61968 & 39.27 & 1.15 & 6.78 & MRS \\
 \noalign{\smallskip}\hline
 \end{tabular}
 \end{table}
  \clearpage

\section{{\it isochrones} fitting results}
\label{iso_appendix.sec}
Here we present the {\it isochrones} fitting results for G6084, G6081, and G6405.

\setcounter{figure}{0}

\begin{figure*}[h]
   \center
   \includegraphics[width=0.98\textwidth]{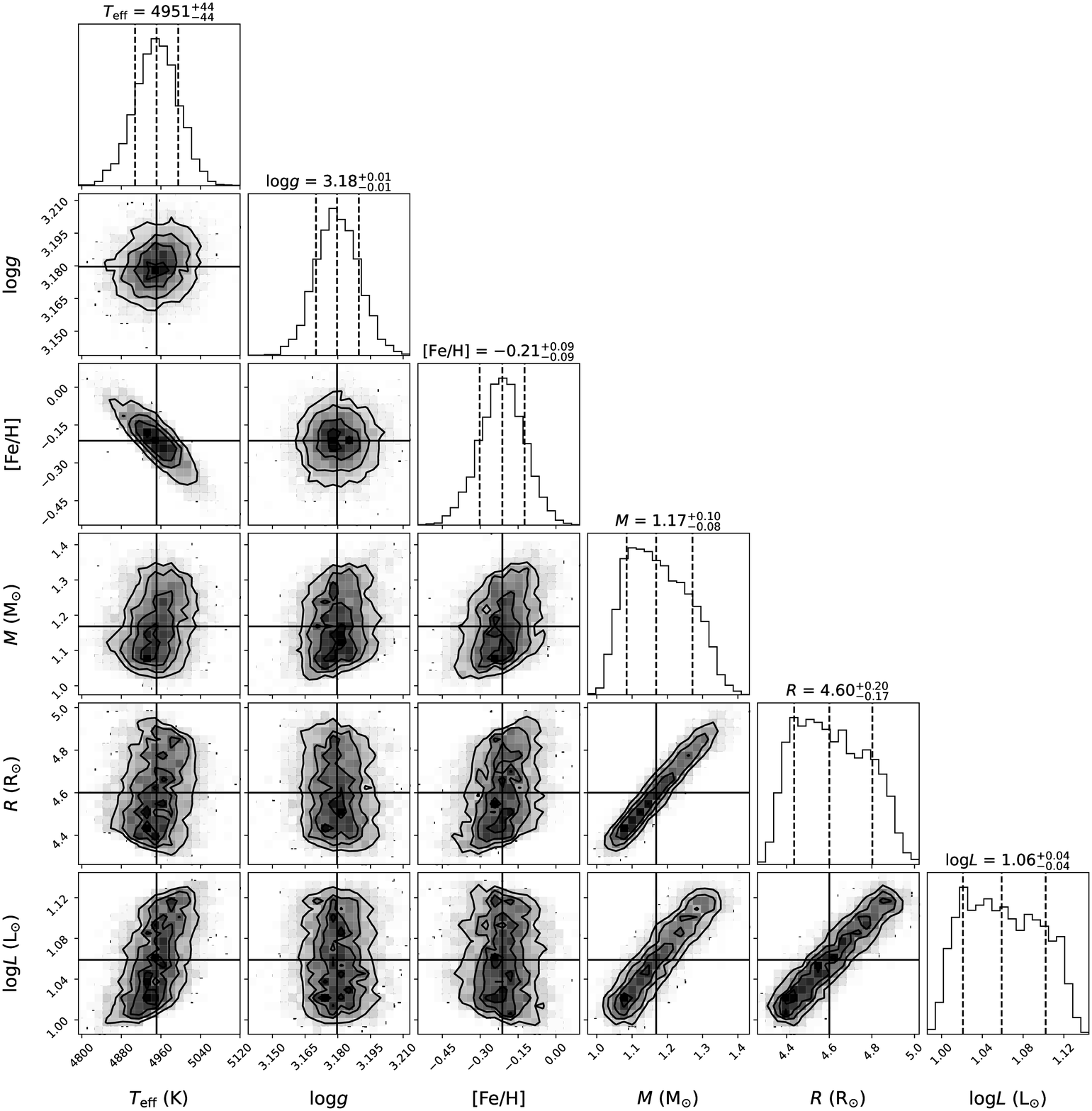}
   \caption{Corner plot showing the distribution of observed and physical parameters of G6084, derived from the {\it isochrones} code. The parameters are labeled as, effective temperature ($T_{\rm eff}$, in K), surface gravity (log$g$, in dex), metallicity ([Fe/H], in dex), mass ($M$, in M$_{\odot}$),  radius ($R$, in R$_{\odot}$), and bolometric luminosity (log$L$, in L$_{\odot}$).}
   \label{8503isochrones.fig}
\end{figure*}

 \begin{figure*}
   \center
   \includegraphics[width=0.98\textwidth]{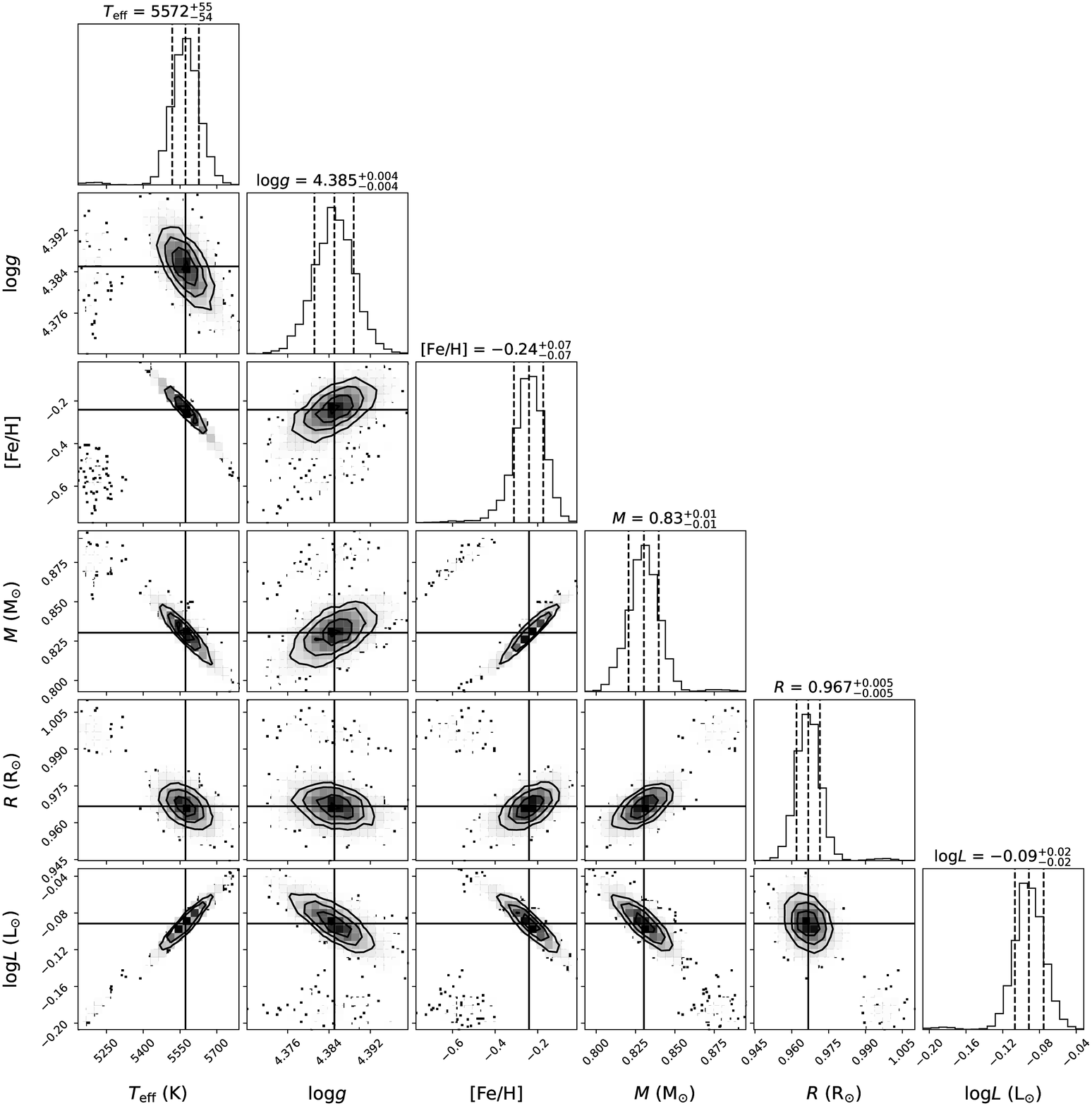}
   \caption{Corner plot showing the distribution of observed and physical parameters of G6081, derived from the {\it isochrones} code. The parameters are as Figure \ref{8503isochrones.fig}.}
   \label{8050isochrones.fig}
 \end{figure*}

\begin{figure*}
   \center
   \includegraphics[width=0.98\textwidth]{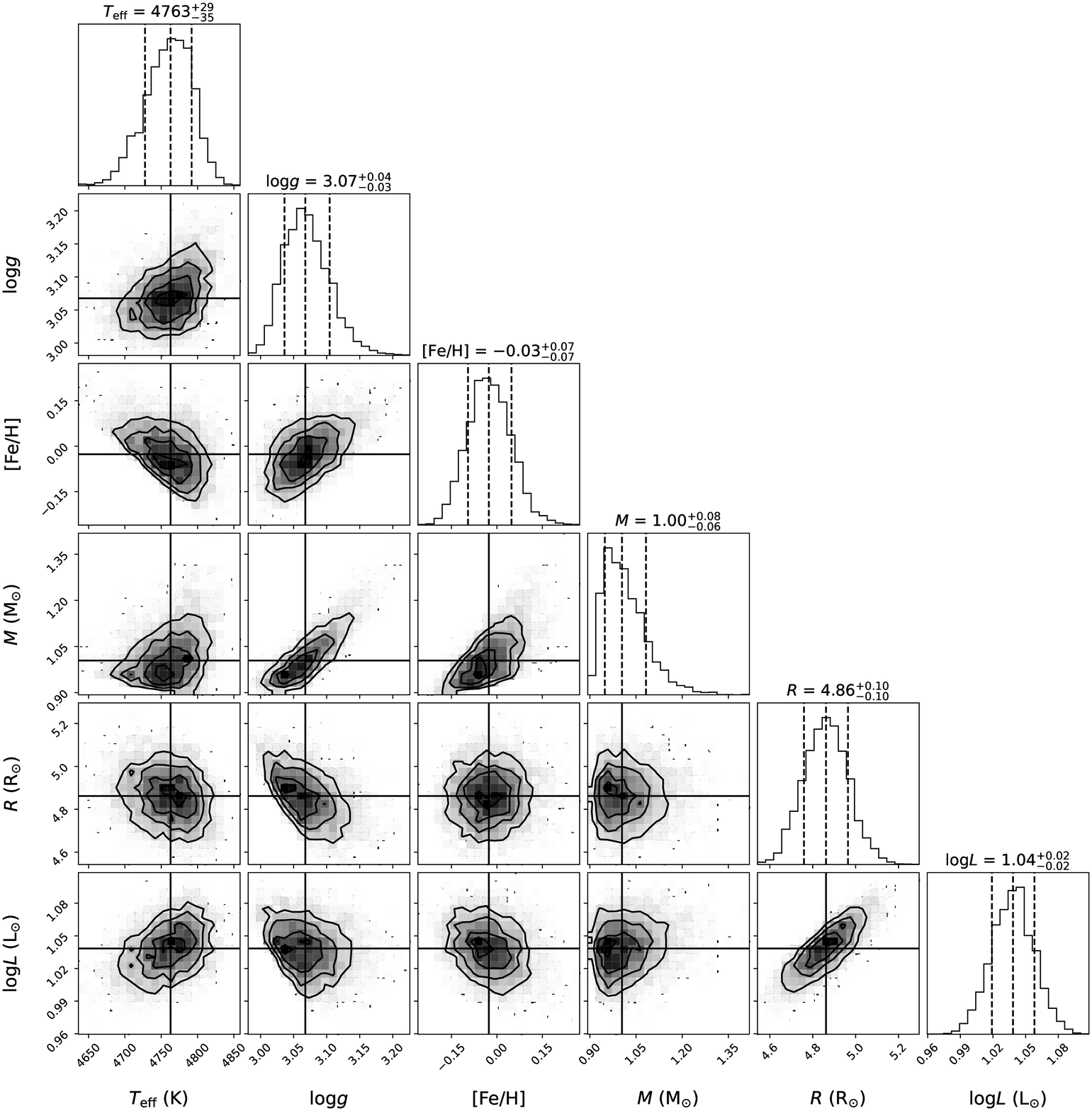}
   \caption{Corner plot showing the distribution of observed and physical parameters of G6405, derived from the {\it isochrones} code, as an example. The parameters are as Figure \ref{8503isochrones.fig}.}
   \label{128isochrones.fig}
\end{figure*}
 \clearpage

\section{{\it The Joker} fitting results.}
\label{joker_appendix.sec}

Here we present the {\it The Joker} MCMC results for G6084, G6081, and G6405.
\setcounter{figure}{0}
\begin{figure*}[h]                                                        
\center 
\includegraphics[width=0.98\textwidth]{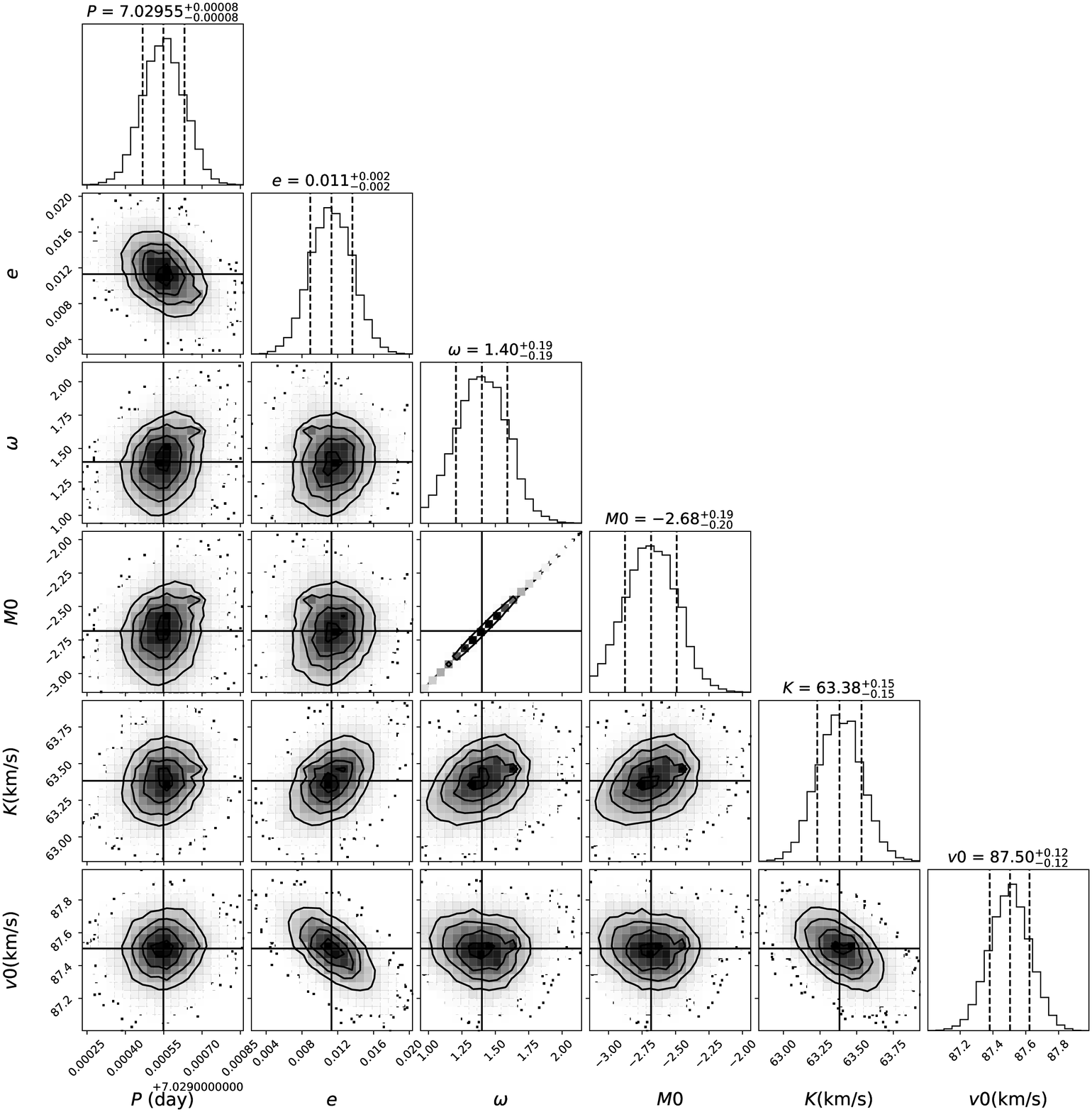}
\caption{Corner plot showing distribution of orbital parameters of G6084, derived from {\it The Joker}. The parameters are labeled as, orbital period ($P$, in days),  eccentricity of the system ($e$), argument of pericenter ($\omega$, in radians), mean anomaly at reference time ($M_{\rm 0}$, in radians), RV semi-amplitude of the star ($K$, in km/s), and the center of mass velocity ($\nu$0, in km/s).}
\label{8503jokermcmc.fig}
\end{figure*}

\begin{figure*}  
\center 
\includegraphics[width=0.98\textwidth]{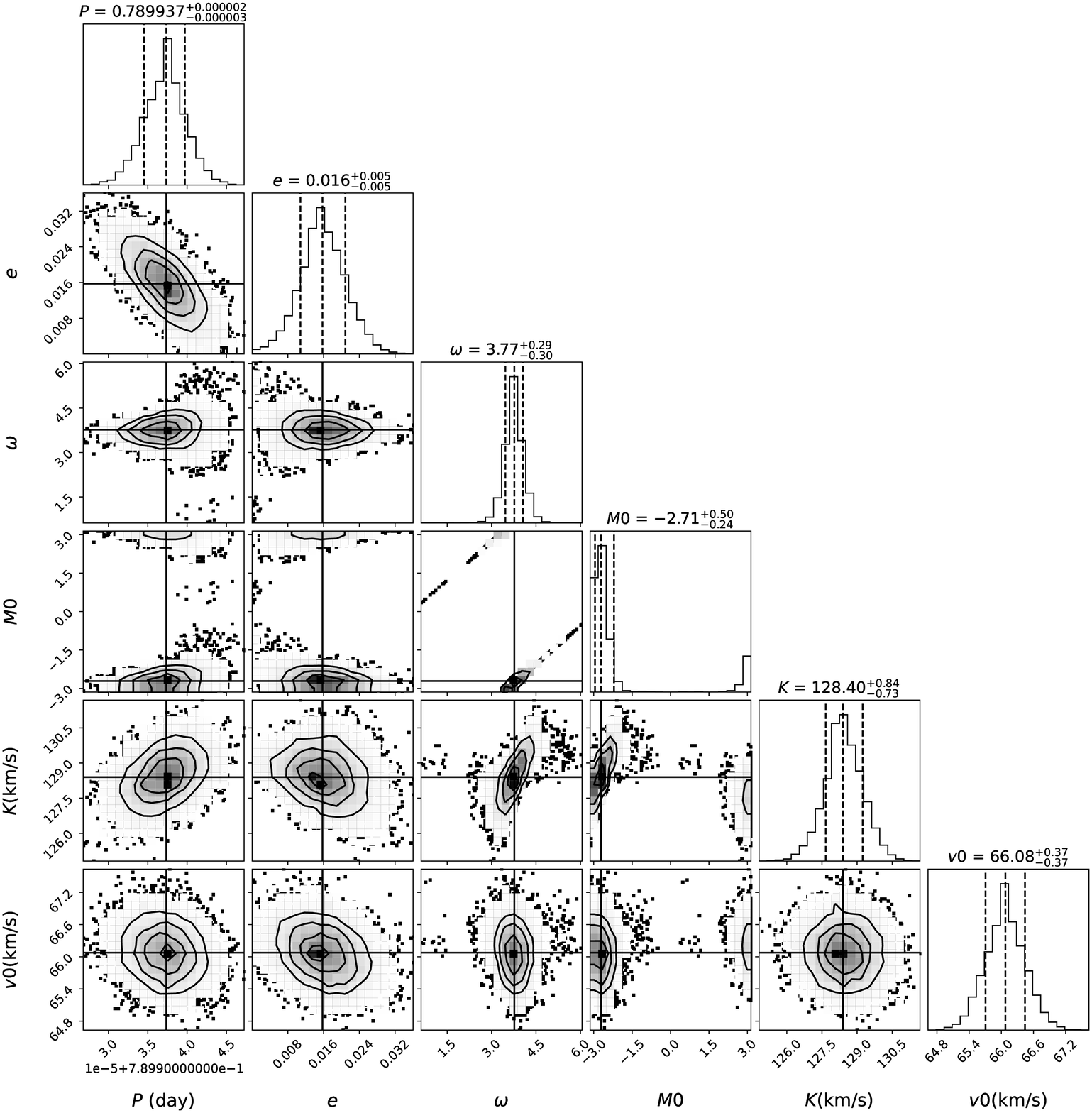}
\caption{Corner plot showing distribution of orbital parameters of G6081, derived from {\it The Joker}. The parameters are as Figure \ref{8503jokermcmc.fig}.}
\label{8050jokermcmc.fig}
\end{figure*}

\begin{figure*}                                                          
\center 
\includegraphics[width=0.98\textwidth]{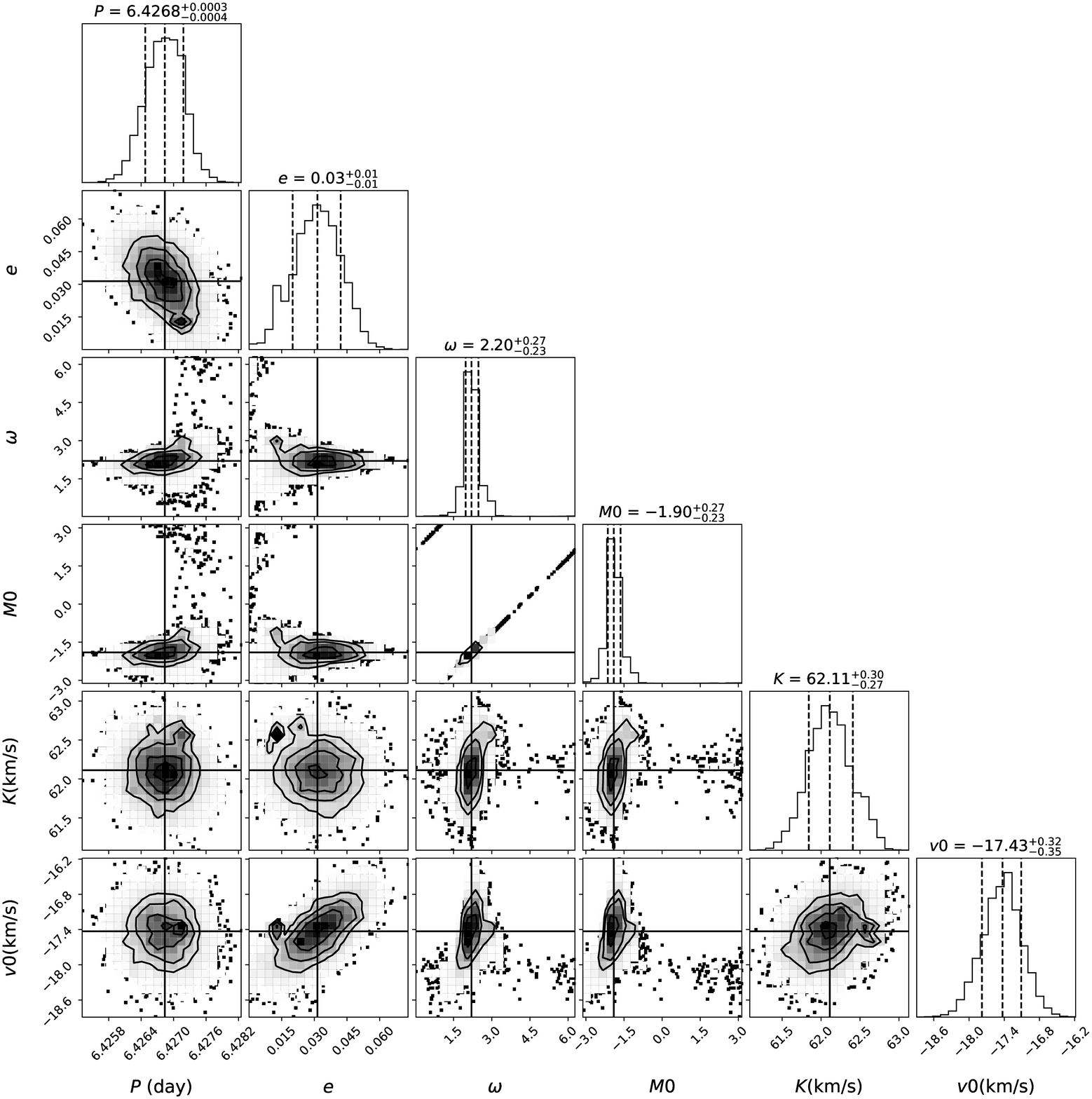}
\caption{Corner plot showing distribution of orbital parameters of G6405, derived from {\it The Joker}. The parameters are as Figure \ref{8503jokermcmc.fig}.}
\label{128jokermcmc.fig}
\end{figure*}

\end{document}